\documentclass[a4paper,11pt]{article}
\pdfoutput=1 
\usepackage{jheppub} 

 \usepackage{graphicx}
 \usepackage{subcaption}
\graphicspath{ {./} }                    
\usepackage{slashed}
\usepackage[T1]{fontenc} 
\usepackage{amssymb}
\usepackage{mathrsfs}
\usepackage{float}
\usepackage{hyperref}
\usepackage{cancel}
\usepackage{caption}
\usepackage{subcaption}
\usepackage[normalem]{ulem}
\usepackage{color}

\newcommand{\beq}{\begin{equation}}
\newcommand{\eeq}{\end{equation}}
\def\barr{\begin{array}}
\def\earr{\end{array}}
\def\dis{\displaystyle}
\newcommand{\bsp}{\begin{split}}
\newcommand{\esp}{\end{split}}
\newcommand{\bit}{\begin{itemize}}
\newcommand{\eit}{\end{itemize}}
\definecolor{darkcyan}{cmyk}{1,0,0,0.4}

\title{\boldmath A closed clockwork theory: $\mathbb{Z}_2$ parity and more}

\author{Debajyoti Choudhury}
\author{and Suvam Maharana}
 \affiliation{Department of Physics and Astrophysics, University of Delhi, Delhi 110007, India}

\emailAdd{debchou.physics@gmail.com}
\emailAdd{smaharana@physics.du.ac.in}

\abstract{We develop a new class of clockwork theories with an
  augmented structure of the near-neighbour interactions along a
  one-dimensional \textit{closed} chain. Such a topology leads to new
  and attractive features in addition to generating light states with
  hierarchical couplings via the usual clockwork mechanism. For one,
  there emerges a $\mathbb{Z}_2$ symmetry under the exchange of fields
  resulting in a physical spectrum consisting of states, respectively
  even and odd under the \textit{exchange} parity with a two-fold
  degeneracy at each level. The lightest odd particle, being
  absolutely stable, could be envisaged as a potential dark matter
  candidate. The theory can also be obtained as a deconstruction of a
  five-dimensional theory embedded in a geometry generated by a linear
  dilaton theory on a $S^1/\mathbb{Z}_2$ orbifold with three
  equidistant 3-branes. Analogous to the discrete picture, the
  $\mathbb{Z}_2$ symmetry in the bulk theory necessitates the
  existence of a KK spectrum of even and odd states, with doubly
  degenerate modes at each KK level when subject to certain boundary
  conditions.}

\begin{document} 
\maketitle
\flushbottom

\section{Introduction}
\label{sec:intro}
As successful as it might seem, the Standard Model (SM) of particle
physics is, by no means, the ultimate theory 
describing Nature in its totality.  The shortcomings, both
theoretical and those pertaining to discrepancies with data have led
to continuing searches for realistic extensions.  Over the years, a
plethora of phenomenologically relevant new physics models have been
proposed to address the shortcomings of the
SM, but only for newer issues to emerge.  A recurrent problem pertains
to the apparent hierarchy of mass scales and couplings. For instance,
models with axions or axion-like particles typically require
small axion-SM couplings \cite{Kawasaki:2013ae}. The case of neutrino
mass models with hierarchically heavy right-handed neutrinos serves as
another example \cite{King:2003jb,deGouvea:2016qpx}. Even within the
SM, the fermion masses display a hierarchy among the different
flavors. The issue, therefore, warrants an underlying mechanism to
generate such small masses or couplings, the absence of which would
require a fine-tuning in the corresponding ultraviolet (UV)-complete
theory.

The \textit{clockwork} mechanism proposed in
refs.\cite{Kaplan:2015fuy,Choi:2015fiu,Giudice:2016yja} apparently
offers an interesting solution to the aforementioned conundrum. The
original mechanism (applicable to particles of any given spin),
generically, entails $N+1$ copies of a field theory, each involving a
massless field with the masslessness being protected by some global
symmetry $G$. The entire construction can be pictured as a chain of
sites in a one-dimensional lattice in the field space with the full
symmetry being $G^{N+1}$. Furthermore, introducing near-neighbour
interaction terms with a strength parametrised by a dimensionless
factor $ q \, (>1)$ explicitly breaks the global symmetry to a single
factor $G'$ corresponding to which we obtain a single massless
particle.  This massless state, a linear combination of the $N+1$
fields, interestingly, has a hierarchical distribution along the sites
with an exponential localisation towards one of the boundary
sites\footnote{While similar effects had been anticipated in
  refs.\cite{Georgi:1989ic,Choudhury:1991if} the goals of (and, hence,
  the field assignments in) the clockwork scenarios are distinctly
  different.}. Consequently, a coupling of the SM sector to the
opposite boundary of the \textit{clockwork} space would have a
coupling of magnitude $\sim 1/q^N$ with the massless state.  Thus,
even without introducing any small coupling, just a
moderately large value of $N$ generates a very sizable suppression.
The massive modes (the so-called clockwork ``gears''), on the other
hand, have nearly-flat distributions along the sites. This is the
essence of the \textit{clockwork} mechanism which has found its
applications in not only the original motivations around axion physics
\cite{Kaplan:2015fuy,Choi:2015fiu,Coy:2017yex,Agrawal:2017cmd,Long:2018nsl,Choi:2014rja,Flacke:2016szy,Davidi:2018sii}
but also in other contexts like flavor hierarchies in masses and
mixings
\cite{Patel:2017pct,Alonso:2018bcg,Babu:2020tnf,vonGersdorff:2020ods},
neutrino masses
\cite{Ibarra:2017tju,Banerjee:2018grm,Kitabayashi:2019qvi}, dark
matter
\cite{Hambye:2016qkf,Kim:2017mtc,Kim:2018xsp,Goudelis:2018xqi,Bernal:2020yqg},
neutron-antineutron oscillation \cite{Arun:2022eqs}
as well as inflationary cosmology
\cite{Kehagias:2016kzt,Im:2017eju,Joshipura:2020ibd}. The mechanism,
in its present form, largely accommodates only Abelian groups, but discussions pertaining to non-Abelian clockwork theories
can be found in refs. \cite{Craig:2017cda,Ahmed:2016viu,Kang:2020cxo}.

The formulation of the clockwork theories as described in
ref.\cite{Giudice:2016yja} has another compelling attribute in that the
one-dimensional lattice in the theory space can be regarded as the
deconstruction of a continuum 5D field theory. The corresponding
\textit{warped} geometry turns out to be the one that is generated by
a 5D linear dilaton theory of gravity
\cite{Antoniadis:2011qw,Cox:2012ee} with the fifth-dimension
compactified on a $S^1/\mathbb{Z}_2$ orbifold augmented by a 3-brane
at each of its boundaries. This setup, similar to the RS1 model
\cite{Randall:1999ee}, also offers a possible solution to the
hierarchy problem along with a rich phenomenology of the dynamical
graviton and dilaton modes \cite{2014,2018}.

Since the mechanism relies on the specific structure of the
near-neighbour interactions, it is worth exploring whether there exist
extensions of this structure, with different \textit{topologies}, that
possess novel phenomenological implications in addition to a similar
\textit{clockwork} mechanism. To this effect, we attempt to construct
a class of clockwork theories with a modified structure compared to
the original such that the resulting hierarchy of the massless
particle's distribution along the sites first increases (decreases)
from one end towards an intermediate site $p$ and then decreases
(increases) towards the other end. Further, taking $p$ to be the
centre-most site in the lattice and identifying the two end-sites lead
to a theory with a $\mathbb{Z}_2$ symmetry characterized by an
exchange of fields in the two arms about the $p$-th site.  Another
distinguishing feature of the construction is that it also generates a
spectrum of massive particles (CW gears) which are doubly degenerate
at each level with the exception of the heaviest mode\footnote{This is
  in stark contrast to the {\em two-sided} theory considered in
  ref. \cite{Banerjee:2018grm} in the context of neutrino masses.}.
This is but a consequence of the \textit{exchange} parity that emerges
naturally and ensures that the lightest odd mode in the spectrum
remains absolutely stable and, therefore, could potentially be a dark
matter candidate.

We further identify the continuum counterpart of the new theory with
the geometry generated by a 5D linear dilaton (LD), albeit augmented
by a third 3-brane at the centre of the orbifold, with the
$\mathbb{Z}_2$ symmetry being identified with the coordinate
reflection in the fifth dimension about the middle brane. The 4D
spectrum, for a theory with a vanishing bulk mass, consists of a
purely massless mode and a tower of massive KK excitations with a
two-fold degeneracy at each KK level.  The latter, automatically, are
either odd or even under the continuum version of the \textit{exchange
  parity} which can be regarded as a form of KK parity in a warped 5D
scenario, a feature that is typically absent in two-brane warped
scenarios.

The primary goal of this paper is to present the generic structure of
the modified scenario and, therefore, we resort to discussing the
phenomenological implications only in passing, deferring dedicated
studies thereof to future works. The outline of the paper is as
follows.  In section \ref{sec:pre}, we review the original
\textit{one-sided} clockwork theory and introduce our modified setup
in section \ref{sec:disccw}. Here, we limit the discussion to bosonic
field theories with Abelian symmetries.  Section \ref{sec:contgeom}
delineates the continuum limit and the corresponding linear dilaton
theory that generates the modified 5D metric. The nature of bulk
scalar and Abelian vector fields in the new background geometry is
evaluated individually in section \ref{sec:contcw}. In section \ref{sec:ft} we comment on the fine tunings in the linear dilaton theory and, finally, we
conclude in section \ref{sec:conclu}.

\section{Prelude} \label{sec:pre}
We review here, briskly, the basic elements of the original clockwork
mechanism proposed in refs.\cite{Kaplan:2015fuy,Giudice:2016yja},
specifically the case of the scalar clockwork. The idea essentially
entails a low energy theory of $(N+1)$ Nambu-Goldstone bosons
corresponding to the spontaneous breaking of an Abelian theory with a
global symmetry $U(1)^{N+1}$ at some high scale $f$. The theory below
the scale $f$, therefore, has a Goldstone symmetry $U(1)^{N+1}$ which
is broken explicitly to a single (shift) symmetry $U(1)_{CW}$ by
introducing interaction terms among ``adjacent'' NGB species, namely
\beq \label{eqn:pre1}
\begin{split}
\mathcal{L}_{CW} &= -\frac{f^2}{2}\sum_{j=0}^{N}\partial_\mu U^{\dagger}_{j} \partial^\mu U_{j}-\frac{m^{2}f^{2}}{2}\Bigg[\sum_{j=0}^{N-1}U_{j}^{\dagger}U_{j+1}^{q} \Bigg] + \mbox{h.c.}
\end{split}
\eeq
where $m$ is a mass parameter and $q$ is a real constant. We adopt the metric signature $(-,+,+,+,+)$.  Denoting
the charge operator corresponding to $U(1)_j$ by $Q_j$, (a simple
choice is $Q_j\{U_j\} =1, \forall j$), the generator corresponding to
$U(1)_{CW}$ is
\beq
\mathcal{Q}_{CW}= \sum_{j=0}^{N} \frac{Q_j}{q^{j}},
\eeq
From a different perspective, each $U_j$ defines a site on a discrete
landscape of scalar field theories and eq.\ref{eqn:pre1} simply
denotes a chain of sites with near-neighbour interactions.
Representing $U_{j}=e^{i \phi_{j}(x)/f} \, (j=0\dots N)$ with the
$\phi_j$'s being the pNGBs, the first term in eq.\ref{eqn:pre1}
leads to canonical kinetic terms for the $\phi_j$ while the second
yields a $(N+1)\times(N+1)$ (mass)$^2$ matrix apart from quartic and
higher order terms.

Diagonalizing the free Lagrangian through the unitary transformation
\beq \label{eqn:transbasis}
\varphi_n=\sum_{j} a_{nj}\phi_j, \eeq
we expect one exactly massless mode (on account of the residual shift
symmetry) and $N$ massive modes.  Indeed, the eigenvalues are
\beq \label{eqn:pevscal}
m_{n}^{2}=m^{2}
\left\{ \barr{lcl}
   0 & \qquad & n=0 \\[2ex]
  \dis 1+q^{2}-2q\cos\frac{n\pi}{N+1}& & n \neq 0
  \earr
  \right.
\eeq
with the corresponding eigenvectors being $(j=0 \dots N, \quad n=1\ldots N)$
\beq \label{eqn:pevcscal1}
\barr{rclcrcl}
a_{0j}&=& \dis \frac{\mathcal{N}_{0}}
{q^j}, &\qquad& a_{nj} &= & \dis \mathcal{N}_{n}\left[ q \sin\frac{j n\pi}{N+1} - \sin\frac{(j+1)n\pi}{N+1}\right]  \\[2ex]
\mathcal{N}^2_{0}&\equiv& \dis \frac{q^2-1}{q^2-q^{-2N}}
& &  \mathcal{N}_{n}^2 &\equiv& \dis \frac{2 m^2}{(N+1)m^2_n}.
\earr
\eeq
The \textit{clockwork}(CW) mechanism can now be illustrated with a
simple example. Consider an external operator ({\em i.e.}, composed of
fields other than $\phi_j$) $\mathcal{O}_{\rm ext}$ which couples to
the clockwork sector at the $N$-th site as
\beq
\mathcal{L}_{int}= y \phi_N \mathcal{O}_{\rm ext} \ ,
\eeq
where $y$ is a coupling parameter. In the mass basis, this results in
\beq
\mathcal{L}_{int}= y \left[\frac{\mathcal{N}_{0}}
{q^N}\varphi_{0} + \sum_{n=1}^{N}\mathcal{N}_{n}q \sin\frac{N n\pi}{N+1}
\varphi_{n} \right] \, \mathcal{O}_{\rm ext} .
\eeq
Assuming that all the dimensionless parameters in the theory are
\textit{a priori} $\mathcal{O}(1)$, we see that $\varphi_0$ couples to
the external operator with an exponentially suppressed strength for
$q>1$ and a sufficiently large $N$. On the other hand, the heavy
eigenstates couple to the external operator with a nearly flat
coupling profile. Thus, the clockwork mechanism generates very small
couplings naturally in a theory which does not contain any small
parameters whatsoever. This is the crux of the mechanism in a
nutshell. In the modified theory that we discuss in the subsequent
sections we largely retain this feature, albeit with small
modifications.  In addition, several new and attractive features
emanate automatically.

\section{The closed clockwork} \label{sec:disccw}
As seen in the preceding section, the original clockwork scenario
could be thought of as an open chain of $U_j$'s with their charges,
under the remnant $U(1)_{CW}$, being given by a geometric progression.
In contrast, we consider a theory with a closed chain, restricting
ourselves to a discussion of only scalars and Abelian gauge
fields\footnote{Incorporating non-Abelian groups in discrete clockwork
  scenarios introduces certain complications. For example, in the
  scalar theory, if the fields $U_j$ were to be bi-fundamental
  representations of global non-Abelian groups $G_{j,L}\times
  G_{j,R}$, the non-trivial case $q>1$ preserves (post explicit
  breaking of the full symmetry) only a diagonal subgroup
  $H$. Consequently, this leads to a pseudo-Goldstone multiplet
  susceptible to loop-level corrections to its mass
  \cite{Ahmed:2016viu}. On the other hand, non-Abelian gauge
  invariance would typically necessitate the trivial case $q=1$ for
  which the clockwork mechanism does not exist. One may, however,
  choose to invoke a clockwork hierarchy at the expense of losing
  gauge invariance \cite{Craig:2017cda,Giudice:2017suc}.}.

\subsection{Scalar Clockwork} \label{subsec:discscal}
A closed clockwork chain implies, of course,
that either the charge ratio $q$ has to be unity or that the geometric progression needs
to invert at a point. Since the first alternative does not lead
  to a hierarchy of couplings, we adopt the second. To be very
specific, we consider an even number $2p$ of pseudo-NGBs and modify
the Lagrangian of eq.\ref{eqn:pre1} to
\beq \label{eqn:lscal2}
\begin{split}
\mathcal{L}_{CW}&= -\frac{f^2}{2}\sum_{j=0}^{2p-1}\partial_\mu U^{\dagger}_{j} \partial^\mu U_{j} -\frac{m^{2}f^{2}}{2}\Bigg[ \sum_{j=0}^{p-1}U_{j}^{\dagger q}U_{j+1} \, + \, \sum_{j=p}^{2p-1}U_{j}^{\dagger}U_{j+1}^{q} \Bigg] + \mbox{h.c.} \\
&=-\frac{1}{2}\sum_{j=0}^{2p-1}\partial^\mu \phi_{j} \partial_\mu \phi_{j}+\frac{m^{2}}{2}\Bigg[ \sum_{j=0}^{p-1}(q\phi_{j}-\phi_{j+1})^{2} \, + \, \sum_{j=p}^{2p-1}(\phi_{j}-q\phi_{j+1})^{2} \Bigg] + \textit{O}(\phi^{4}) \ .
\end{split}
\eeq
Note that we have maintained the same charge ratio $q$ in both arms
  which necessitates the pivoting to occur exactly midway, {\em i.e.}, at the $p^{\rm th}$ site. Unequal ratios in the two arms could be chosen too, but only at the cost of losing some very attractive features.

The identification $U_{2p} \equiv U_0$ (ensuring that
$\phi_{0}$ couples to both $\phi_1$ and $\phi_{2p-1}$) renders this a
\textit{closed
  clockwork} scenario and distinguishes it from the two-sided but
open-ended scenario presented in ref.~\cite{Banerjee:2018grm}. This single
difference turns out to have profound implications as we would see below.

Before discussing the mass spectrum, it
  is worthwhile to enumerate the symmetries of the theory.  As
before, the Lagrangian in eq.\ref{eqn:lscal2} has a residual shift
symmetry $U(1)_{CW}$ which is now defined by the generator,
\beq \label{eqn:cwgen2}
\mathcal{Q}_{CW}=\sum_{j=0}^{p-1} \frac{Q_j}{q^{2p-j}} + \sum_{j=p}^{2p-1} \frac{Q_j}{q^{j}},
\eeq
where, as before, we have assumed that each $U_j$ has a unit charge
under the symmetry $U(1)_j$. Additionally, the theory has a
$\mathbb{Z}_2 \otimes \mathbb{Z}_2$ symmetry, where the first
$\mathbb{Z}_2$ corresponds to the rigid exchanges of the fields,
namely, $\phi_j \to \phi_{2p-j}$ and the second to the exchanges
$\phi_j \to -\phi_{2p-j}$, both $\forall j$.\footnote{Equivalently,
  these can be recast as the trivial rigid $\phi_j \to - \phi_j$ and
  $\phi_j \to \phi_{2p-j}$, where the former simply reflects the invariance under charge conjugation (namely, $U_j \to U_j^*$).  The second symmetry, therefore, is the only nontrivial discrete transformation under which the theory is invariant.}  A consequence of these symmetries is
that the eigenmodes must be either even or odd under it. In other
words, the eigenstates may be represented
as
\beq \label{eqn:transbasis2}
\varphi^{(\pm)}_n=\sum_{j} a^{(\pm)}_{nj}\phi_j,
\eeq
where $(\pm)$ denote the even and odd modes respectively and $a^{(\pm)}_{nj}$ are the corresponding elements of the transformation matrix, satisfying
the eigenvalue equation,
\beq \label{eqn:eveq}
\sum_j [M_{\phi}^2]_{i j} a_{n j}=m^2\lambda_n a_{n i} \ , \forall n, i\ .
\eeq

\begin{figure*}[tbp] 
\centering
      \includegraphics[scale=0.4,keepaspectratio=true]{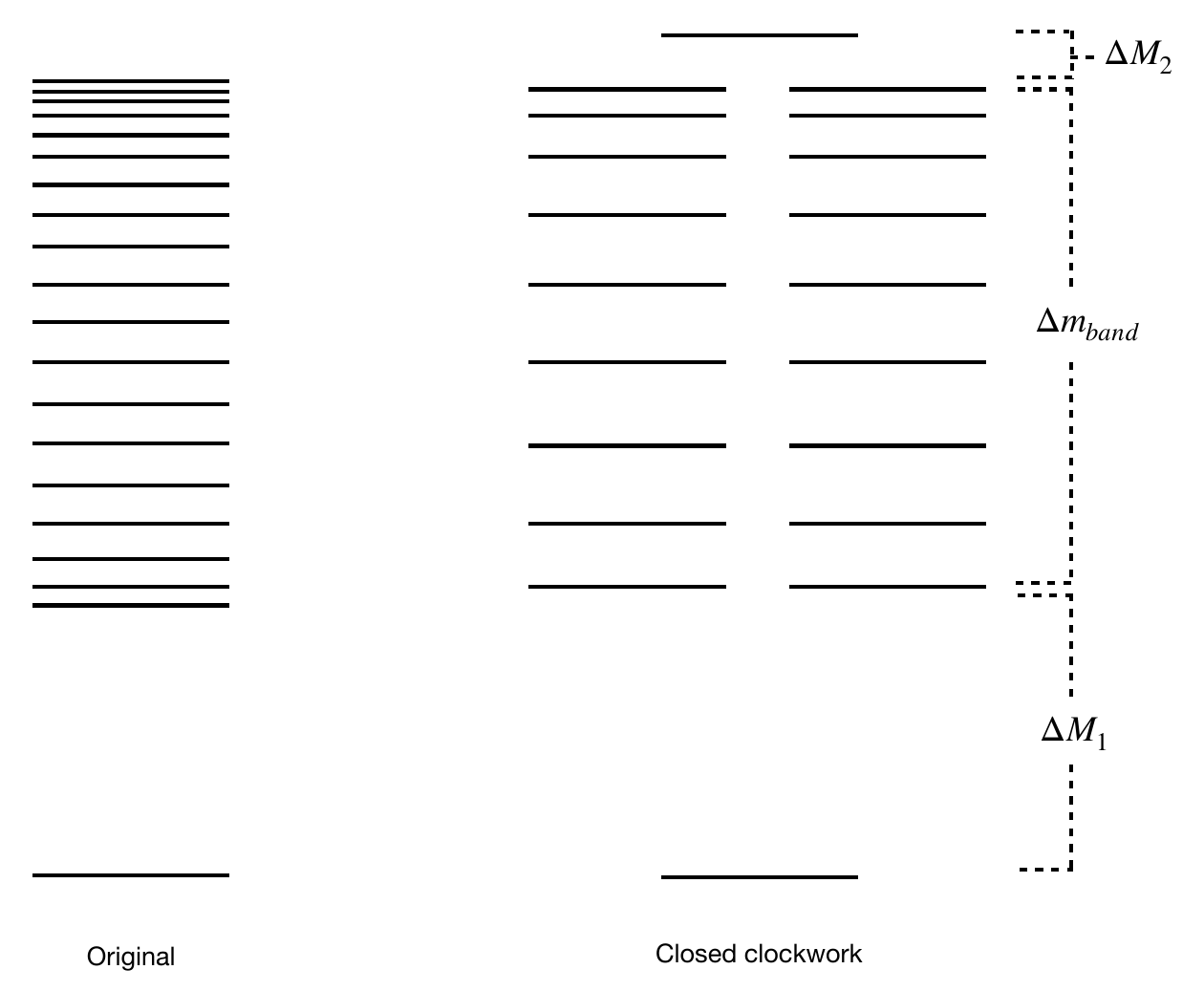}
	\caption{Schematic depiction of the physical states in the \textit{closed} clockwork theory (right) and in the original one-sided theory (left) for $N=20$.} \label{fig:spec_disc}  
\end{figure*}

Reverting to the $(2p\times 2p)$ mass matrix, it
is given by
\beq \label{eqn:mscal}
M_{\phi}^{2}=m^{2}\left(
\begin{array}{cccccccccc}
& \mbox{\tiny (col $0$)} & \quad & \quad & \quad \mbox{\tiny (col $p$)} & \quad & \quad & \quad & \mbox{\tiny (col $2p-1$)}\\
\mbox{\tiny (row $0$)} & 2 q^2 & \quad -q &  \quad 0 & \quad 0 & \quad \ldots & \quad \ldots &\quad \ldots  & \quad -q \\
 &-q & \quad q^2+1 & \quad -q &\quad 0 & \quad \ldots & \quad \ldots &\quad \ldots & \quad 0 \\
 & 0 &\quad \ddots & \quad \ddots &\quad & \quad \ldots & \quad \ldots & \quad \ldots & \quad \vdots\\
 & \vdots & \quad & \quad \ddots & \quad \ddots & \quad \ldots & \quad \ldots & \quad \ldots  & \quad \vdots\\
   
& \ldots &\quad -q &\quad q^2+1 &\quad -q &\quad 0 &\quad \dots &\quad \ldots  &\quad \vdots \\
\mbox{\tiny (row $p$)}& 0 &\quad \dots &\quad -q &\quad 2 &\quad -q &\quad 0 &\quad \ldots  &\quad \ldots\\
 &0 &\quad \ldots &\quad \ldots &\quad -q &\quad q^2+1 &\quad -q &\quad 0 &\quad \vdots \\
 & \vdots &\quad \ldots & \quad \ldots &\quad & \quad \ddots & \quad \ddots & \quad \ldots & \quad \vdots\\
 
\mbox{\tiny (row $2p-1$)}&-q &\quad 0 &\quad 0 &\quad 0 &\quad \ldots &\quad \ldots &\quad -q &\quad q^2+1 \\
\end{array}
\right). 
\eeq
When compared to the mass matrices obtained
in ref.\cite{Giudice:2016yja} from eq.\ref{eqn:pre1} or in
ref.\cite{Banerjee:2018grm}, differences appear only in the
$(0,0), (p, p), (0, 2p-1), (2p-1,0)$ and $(2p-1, 2p-1)$ elements and,
together, these would turn out to have very interesting consequences.
For one, the eigenvalues are
\beq \label{eqn:evscal}
m_{n}^{2}\equiv m^2 \lambda_n=m^{2} \begin{cases}
0 & n=0 \ ,\\
2(1+q^2) & n=p \ ,\\ \dis
1+q^{2}-2q\cos \frac{n\pi}{p} & \mbox{otherwise} \ .
\end{cases}
\eeq
A few remarks on the physical spectrum are now in order
\begin{itemize}
  \item As is evident from eq.\ref{eqn:evscal}, the spectrum
    consists of an exactly massless mode $\varphi_0$ and is bounded
    from above by the massive mode $\varphi_p$.
\item In addition to the mass gap between the lightest mode
  $\varphi_0$ and the level $1$ modes $\varphi_1/\varphi_{2p-1}$
  ($\Delta M_1$) there exists another one ($\Delta
  M_2$) between the heaviest state $\varphi_p$ and the
  penultimate $\varphi_{p-1}/\varphi_{p+1}$ states\footnote{With
      $\Delta M_{1,2}^2 \approx m^2 (q-1)^2$, the second gap ($\Delta M_2$)
      is noticeable only
      for large $q$.}. This is in contrast with the case in
  \cite{Kaplan:2015fuy,Giudice:2016yja} wherein only the lower gap
  exists.
  
\item The intermediate states are doubly degenerate at each level,
  with each degenerate subspace being spanned by the eigenmodes
  $\varphi_n$ and $\varphi_{2p-n}$, or, equivalently, $\varphi^\pm_n$.
  These fill a band of width $\Delta m_{\hbox{\tiny band}} \sim 2m$
  with mass-splittings $\delta m_n/m_n$ of $\mathcal{O}(1/2p)$ for
  sufficiently large $p$. The mass-splitting first increases in the
  band as we move up the spectrum and then decreases progressively
  towards the end of the band. Fig.\ref{fig:spec_disc} shows a
  schematic of the physical modes emerging in the closed clockwork
  model along with that of the original model for comparison. Clearly,
  the inter-level spacings in the closed scenario are larger than
  those seen in the original one-sided clockwork theories.
\item
 This degeneracy is not simply a consequence of the $\mathbb{Z}_2$
 symmetry alone\footnote{Were that to be the case, such a degeneracy
   could as well be seen in a two-sided open chain which, evidently,
   we do not encounter.}. Rather, this symmetry itself is just a consequence of the assumptions undertaken in our
 construction, namely, the universality ({\em i.e.}, the
 site-independence) of the $U(1)_j$ charges $Q_j$ and the mass
 parameters $m^2$ and $f^2$ as well as the \textit{closed} nature of the
 interactions in the sense described before. Together, these lead to a
   pseudo-tridiagonal mass-squared matrix, with a very specified
   distortion. The corresponding eigenvalue equations are nothing but
   second-order $q$-difference equations, resulting in a
   degenerate spectrum. Any deformity in the universal nature of the
 couplings and/or the structure of the interactions near the pivots
 would lead to non-degeneracies in the masses even if the
 $\mathbb{Z}_2$ symmetry exists.
 The details can be found
 in Appendix \ref{subsec:disdeg}.

 It should be noted that, for small deviations
 from uniformity, we would obtain quasi-degenerate states.
\end{itemize}

The discussion above should convince the reader that this setup is not merely a trivial joining
of two open-ended CW chains, but a distinct
theory with unique properties.

The aforementioned transformation
matrix elements $a^{(\pm)}_{nj}$ are given by
\begin{equation} \label{eqn:evcscal1}
a^{(+)}_{0j}=\mathcal{N}^{+}_{0} \begin{cases}
q^{j-2p}, & 0\leq j \leq p \\
q^{-j}, & j>p
\end{cases} ,
\qquad
a^{(+)}_{pj}=\mathcal{N}^{+}_{p} \begin{cases}
(-q)^{-j}, & 0\leq j \leq p \\
(-q)^{j-2p}, & j>p   \, \, 
\end{cases}
\end{equation}

\begin{equation} \label{eqn:evcscal3}
  a_{nj}^{(+)}(n<p)= \mathcal{N}^{+}_{n}
  \left\{
  \barr{lcl}
  \dis \sin \frac{jn \pi}{p} + \frac{2 q}{q^2-1}\,
  \sin \frac{n \pi}{p} \,  \cos \frac{jn \pi}{p},
  & \qquad & 0\leq j\leq p \\[2ex]
  \dis
  -\sin \frac{jn \pi}{p} +\frac{2q}{q^2-1}
  \sin \frac{n \pi}{p} \, \cos \frac{jn \pi}{p},
  & & \dis p < j\leq 2p-1 
  \earr
  \right.
\end{equation}
and,
\begin{equation} \label{eqn:evcscal4}
a_{nj}^{(-)}(n>p)= \mathcal{N}^{-}_{n}\\
      \sin{\frac{jn \pi}{p}}.
\end{equation}
The normalization factors $\mathcal{N}_{0,p}$ and $\mathcal{N}^{\pm}_{n}$
are given by 
\begin{equation} \label{eqn:normscal}
\barr{rclcrcl}
\mathcal{N}_{0}^2& = & \dis \frac{q^{4p}(q^2-1)}{(q^{2p}-1)(q^{2}+1)}
  & \qquad & 
  \mathcal{N}^2_{p}& = & \dis
  \frac{q^{2p}(q^2-1)}{(q^{2p}-1)(q^{2}+1)}
    \\[2.5ex]
    \dis \left(\mathcal{N}^{+}_{n}\right)^2 & = & \dis
    \frac{(q^2-1)^2 }{p \left[1+q^4-2q^2 \cos{\frac{2n \pi}{p}}\right]}  & & 
  \dis  \left(\mathcal{N}^{-}_{n}\right)^2& = & \dis p^{-1} \ .
\earr
\end{equation}

 Having established the identity of the physical states, let us
 now consider an external operator $\mathcal{O}_E$ (obviously
 transforming trivially under the $\mathbb{Z}_2$) coupling to the CW
 sector only at the $j$-th site, {\em viz.},
\beq \label{eqn:coupex1}
\mathcal{L}_{\rm ext} = -\mathcal{
  O}_E \phi_j \, = \, -\left[ \sum_{n=0}^{p}a^{+}_{j n} \varphi^{+}_n +
  \sum_{n= p +1}^{2p-1} a^{-}_{j n} \varphi^{-}_n \right] \,
   \mathcal{O}_E \ .
\eeq
The $\mathbb{Z}_2$ symmetry is, thus, explicitly broken\footnote{It
  would have remained unbroken had $\mathcal{O}_E$ coupled with
  identical strength to $\phi_{2p-j}$ as well.} as the external
operator couples to both even and odd modes. Note, though, that if the
coupling site is chosen to be either of $j=0, p$, the operator couples
only to the even modes as the odd wavefunctions vanish there. The
mixing matrix elements in eq.\ref{eqn:evcscal1}-\ref{eqn:evcscal4}
specify that the light mode ($\varphi_0$) is localised towards the
$j=p$ site, away from which its distribution falls exponentially,
acquiring a weight $\sim q^{-p}$ at $j=0$. In contrast, the heaviest
mode ($\varphi_p$) is localised near $j=0$, with its weight decreasing
exponentially towards $j=p$ along both arms. Thus, at $j=0$,
$\mathcal{O}_E$ couples to $\varphi_0$ with a suppressed coupling
($\sim q^{-p}$) and to $\varphi_p$ with a strength $\sim
\mathcal{O}(1)$. This order of strengths reverses for an
$\mathcal{O}_E$ coupling at the $j=p$ site instead. On the other hand,
the rest of the massive modes have a nearly-flat distribution along
the sites.

  \subsubsection{A possible dark matter candidate?} \label{subsubsec: dmc}
  The preceding discussion establishes that, in the event of the SM
  coupling with the CW sector at either of the sites at $j=0$ or
  $j=p$, the theory preserves the exchange parity ({\em i.e.}, the
  $\mathbb{Z}_2$ symmetry). Consequently, the lightest odd CW particle
  (LOCP) is rendered absolutely stable and, hence, can potentially be
  a dark matter (DM) candidate (and, by the same token, lead to
  additional sources of missing-momentum signals at colliders). There
  is, though, a key difference with most popular DM models. The
  vanishing of the $\mathbb{Z}_2$-odd CW state wavefunctions at the
  ($\mathbb{Z}_2$) fixed points implies that these do not couple
  directly to the SM sector.  Consequently, the requisite relic
  abundance would be realised primarily through a pair of DM particles
  annihilating into the even CW modes which can further annihilate
  and/or decay into SM particles. Thus, the odd tower would act as a
  secluded dark sector. In such models, co-annihilations could play a
  significant role in enhancing the total cross-sections by virtue of
  the degeneracy in the physical states. Of course, the extent of such
  enhancements would depend on the mass-splittings dictated by the
  parameters of the theory. Another mode of enhancement in the
  annihilation could be effected by introducing couplings that break
  $U(1)_{CW}$ softly, e.g., a trilinear coupling of CW fields that
  opens a new channel of pair annihilation and co-annihilation via a
  mediator which could further introduce resonance enhancements in
  $s$-channel processes and/or enhancements from the lightness of the
  mediator in $t$-channel processes. Typically, for a small number of
  sites ($p \lesssim 5$), the latter
  channel would prove to be the dominant one. We limit the discussion
  here to only highlighting the possibility of the aforementioned DM
  phenomenology and leave the detailed analysis to an accompanying
  work.

\subsection{Vector Clockwork} \label{subsec:disvec}
Having defined the scalar theory, one for vector fields
proceeds quite analogously.  We start with a $U(1)^{2p}$ theory with
identical gauge couplings $g$ alongwith $2p$ copies of complex scalar
fields $\phi_j$, such that the Lagrangian is
\beq \label{eq:vectcw}
\mathcal{L}_{CW}=\sum _{j=0}^{2p-1}\left\{ -\frac {1} {4}F_{\mu
  \nu j} F_{j}^{\mu \nu }-\left( D_{\mu }\phi _{j}\right) ^{\ast
}\left( D^{\mu }\phi _{j}\right) -\lambda \left( \left| \phi
_{j}\right| ^{2}-\frac {f^{2}} {2}\right) ^{2}\right\},
\eeq
with the identification $(F_{N}, \phi_{N})\equiv (F_{0}, \phi_{0})$.
Although the gauge symmetry does not require\footnote{Similarly, the
  terms $\sum_{i,j}|\phi_i|^2|\phi_j|^2$ are also allowed by symmetry
  considerations. However, since their introduction does not bring
  about a qualitative change, we choose to omit them for the sake of
  simplicity.}  the parameters $\lambda$ and $f$ to be
site-independent, we, nonetheless, assume universal values.  This
restriction not only leads to a mass matrix similar to that obtained
in the scalar case with the attendant phenomenology (see the
discussion in section \ref{subsec:discscal} and
Appendix \ref{subsec:disdeg}), but also facilitates an
embedding in a higher dimensional theory. The complex scalars
$\phi_{j}$ have non-zero charges under $U(1)_{j}\times U(1)_{j+1}$,
{\em viz.}
  \beq
\begin{split}
\left( j=0,\ldots,p-1\right) &:\left( q, \, -1\right) \nonumber \\
\left( j=p,\ldots,N-1\right) &:\left( -1, \, q\right) \ .
\end{split}
\eeq

With this configuration in place, an inspection of the symmetries in
eq.\ref{eq:vectcw} shows that each of the complex fields
$\phi_j$ is uncharged under a particular combination
of the Abelian symmetry transformations, namely the one
corresponding to the generator $\mathcal{Q}_{CW}$ given in
eq.\ref{eqn:cwgen2}. This implies that, even on {\em all} the scalars $\phi_j$ acquiring non-zero vacuum
  expectation values, one of the $2p$ $U(1)$ factors must remain unbroken,
  and the
  physical spectrum would include a massless gauge boson and a massless
NGB. Following the SSB, the complex scalars can be represented as,
\begin{equation}
\phi_j=\frac{1}{\sqrt{2}}\left( \chi_j + f\right)e^{i \pi_j/f},
\end{equation}
where $\chi_j$ and $\pi_j$ are real. Integrating out the
heavy radial degrees of freedom $\chi_j$, we have a theory of massive
$U(1)$ gauge fields below the SSB scale $f$ defined by
\beq \label{eq:lvec2}
\barr{rcl}
\mathcal{L}_{\rm eff} =  
- \dis \sum _{j=0}^{2p-1}\frac {1} {4}\left( F_{\mu j}\right) ^{2}
&+& \dis\frac {g^{2}f^{2}} {2} \Bigg\{\sum _{j=0}^{p-1}\left[ \frac{1}{f}\partial_\mu \pi_j +g \left(qA_{\mu j}-A_{\mu j+1}\right)\right]^{2} \\[3ex]
&+& \dis \sum _{j=p}^{2p-1}\left[\frac{1}{f}\partial_\mu \pi_j + g \left( A_{\mu j}-qA_{\mu j+1} \right)\right]^{2}\Bigg\} \ .
\earr
\eeq 
apart from a gauge-fixing term, that we specify later.
As eq.\ref{eq:lvec2}
shows, the mass terms for the
gauge fields resemble those for the scalar CW. In the physical basis, the full theory is now given by,
\beq
\mathcal{L}_{\rm eff}^{\rm (mass)}=\sum_{n=0}^{ 2p-1}\left\{ -\frac {1} {4}\left(
\mathcal{F}_{\mu \nu}^{n}\right)^{2}+\frac {1} {2}m_{n}^{2}\left(
\mathcal{A}_{\mu}^{n} + \frac{1}{m_n} \partial_\mu \tilde{\pi}_{n}\right)^{2}\right\} + \mathcal{L}_{GF}  \ ,
\eeq
where
\beq \label{eqn:mvec}
m_{n}^{2}= \left\{ \barr{lcl}
0 & \qquad & n = 0 \\[1ex]
2 g^{2}f^{2} (1 + q^2) & & n = p \\[1ex]
\dis g^{2}f^{2} \left( 1+q^{2}-2q\cos \frac {n\pi} {p} \right) & & n\neq 0,  p
\earr
\right.
\eeq
 and the eigenstates $\mathcal{A}^{n}_{\mu}$ (corresponding to
$\mathcal{F}$) and $\Tilde{\pi}_{n}$ are defined as
\beq
\barr{lcl}
&\mathcal{A}_{\mu n}& \dis \equiv \sum _{j=0}^{2p-1}a_{n j}A_{\mu j}
\\[6ex]
{\rm and}  \quad &\tilde{\pi}_{n}& \dis \equiv \sum_{j=0}^{2p-1}  b_{nj}\pi_j \\[3ex]
& &= \left\{ \barr{lcl}
\dis \sum_{j=0}^{2p-1} a_{0j} \pi_j & \qquad & n = 0 \\[1ex]
\dis \frac{g f}{m_n}\left[\sum_{j=0}^{p-1}(q a_{nj}- a_{n, j+1})\pi_j + \sum_{j=p}^{2p-1}(-a_{nj}+qa_{n, j+1})\pi_j \right] & \qquad & n > 0 
\earr
\right. \ .
\earr
\eeq
A convenient choice for the gauge-fixing term is given by

\beq \mathcal{L}_{GF} = \frac {-1} {2\xi }
\sum_{n=1}^{2p-1}\left(\partial^{\mu}\mathcal{A}_{\mu n}-\xi m_n
\tilde{\pi}_n \right)^{2 }.  \eeq In the unitary gauge ($\xi \to
\infty$), all of the $(2p-1)$ NGBs decouple from the
theory\footnote{The whole point of choosing the unitary gauge here is
  to elucidate the existence of a massless scalar mode (a Goldstone boson that is not Higgsed) in
    the theory. One could, in general, choose a more convenient gauge
  to facilitate computational simplicity.}. One massless scalar mode
$\Tilde{\pi}_0$, however, remains as a physical state in the spectrum,
as expected from the preceding symmetry argument\footnote{Similar
  entities have been identified before in the context of periodic
  moose structures in
  refs\cite{Cheng:2001vd,Arkani-Hamed:2001nha}}. In summary,
therefore, there are $2p-1$ massive photons with a degeneracy
identical to that of the CW scalars, a massless photon and a massless
scalar. The light scalar may acquire a mass if any of the shift
symmetries is broken explicitly.

Similar to the scalar case in section (\ref{subsec:discscal}) the
complete spectrum obtained here can be classified as even and odd
eigenstates of clockwork parity. Explicitly, the matrix elements
$a_{nj}$ are the same as obtained in
eq.\ref{eqn:evcscal1}--\ref{eqn:normscal}. Consequently, the physical
modes have the same nature of localised couplings with an external
sector as that of the scalar CW in eq.\ref{eqn:coupex1}. The
distinction in this case is that the external operator coupling to the
$j$-th site is a current charged under $U(1)_j$ of the CW sector.

\section{The continuum perspective} \label{sec:contgeom}
Keeping the discrete theories in perspective, we now explore the
possibility of obtaining them as deconstructions of five
dimensional bulk theories, an advantage being that in doing
this, we would encounter some new possibilities. To identify the
appropriate underlying 5D geometry, it is convenient to assume a form
of the metric that incorporates at least those symmetries encountered
in the clockwork theories along the discrete lattice. Since, in our
case, the latter includes a $\mathbb{Z}_2$ symmetry under field
exchanges, a logical ansatz for the 5D metric (with the fifth
dimension being compactified to a circle of radius $R$) would be
\beq
ds^2= X(|z-z_p|)dx^2+Z(|z-z_p|)dz^2,
\eeq
where $X$ and $Z$ are, {\em a priori}, unknown functions of the extra
dimension\footnote{The function $Z$ could, of course, be
    eliminated by a simple redefinition of the coordinate $z$, but it
    is convenient to retain it.} and, by construction, are
$\mathbb{Z}_2$ symmetric about $z_p=\pi R/2$. Here, $x$ denotes the 4D
spacetime coordinates and $z$ is the fifth coordinate. The action for a
massless bulk scalar in this background geometry would, then, be given
by
\beq \label{eqn:bscalaction}
\begin{split}
\mathcal{S}&= \int d^4x \, \int_0^{\pi R}dz \sqrt{-g} \left[ -\frac{1}{2} g^{MN} \partial_{M} \phi(x,z) \, \partial_{N} \phi(x,z)\right] \\
&= -\frac{1}{2} \int d^4x \, \int_0^{\pi R}dz \,  X^{2}Z^{1/2}\left[ X^{-1}(\partial_\mu \phi)^2 + Z^{-1} (\partial_z \phi)^2\right] \ .
\end{split}
\eeq
Re-scaling the scalar field $\phi \to X^{-1/2}Z^{-1/4} \phi$ in order
to obtain canonical kinetic terms post discretization, we have
\beq \label{eqn:bscalaction2}
\begin{split}
\mathcal{S}&= -\frac{1}{2} \int d^4x \, \int_0^{\pi R}dz \left\{ (\partial_\mu \phi)^2 + X^2 Z^{-1/2}\left[\partial_z \left( X^{-1/2}Z^{-1/4} \phi \right)\right]^2\right\} 
\end{split}.
\eeq
We, now, discretize the $z$-dimension into a lattice of $2p$ sites
with uniform inter-site spacing $a=\pi R/2p$ and specify a site $j$ as
$z \to j a$. The $z$-derivative in the second term of the action is to
be replaced by the difference\footnote{Clearly, the derivative could,
  in principle, be mapped into inequivalent differences, and these
  choices would, potentially, lead to different discrete
  theories. This ambiguity is not unexpected, for it is well known
  that several inequivalent discrete theories could flow into the same
  continuum theory.}
\beq
\partial_z \left( X^{-1/2}Z^{-1/4} \phi \right) \to a\left( X_{j+1}^{-1/2}Z_{j+1}^{-1/4} \phi_{j+1} -X_{j}^{-1/2}Z_{j}^{-1/4} \phi_{j} \right). \nonumber
\eeq
Since the discretized derivative involves fields from adjacent sites,
namely, $\phi_{j+1}$ and $\phi_j$, we adopt a particular discretization prescription, wherein
the factor coupled to the derivative is replaced by
\beq \label{eqn:symm}
X^2 Z^{-1/2} \to \left( X_j Z_j^{-1/4}\right)\left( X_{j+1} Z_{j+1}^{-1/4}\right),
\eeq
which consists of equal powers of the functions $X$ and $Z$ evaluated
at adjacent sites $j$ and $j+1$.  With this prescription we have,
\beq \label{eqn:bscalaction3}
\mathcal{S}= \frac{-1}{2}  \int d^4x \, \sum_{j=0}^{2p-1}   \Bigg\{(\partial_\mu \phi_j)^2 
+ \frac{4p^2}{\pi^2 R^2}\left( X_j^{1/2}Z_{j}^{-1/8}Z_{j+1}^{-3/8} \phi_{j+1}-X_{j+1}^{1/2}Z_{j+1}^{-1/8}Z_{j}^{-3/8} \phi_j\right)^2\Bigg\}.
\eeq
The second term inside the braces is a simple square of a difference
of the form $(f_1[j+1]f_2[j]-f_1[j]f_2[j+1])^2$, and the sum is
explicitly invariant under $j \to 2p-j$, a property that would be
absent without the symmetrization encoded in
eq.\ref{eqn:symm}. In other words, the (natural) prescription of
eq.\ref{eqn:symm} preserves, on discretization, the $\mathbb{Z}_2$
symmetry of the continuum theory\footnote{Note that this
  discretization prescription is categorically different from the one
  adopted in \cite{Giudice:2016yja} which would break the
  $\mathbb{Z}_2$ symmetry if employed in our scenario. This also
  demonstrates the fact that different prescriptions can lead to
  different discrete theories which is reminiscent of what is usually
  encountered in the lattice QCD framework, especially in the context
  of discretizing fermions on a lattice. The distinct low-energy
  theories, however, lead to the same bulk 5D theory in the $N \to
  \infty$ limit. It is worth stressing here that this does not
  necessarily mean that the different discrete theories would also
  correspond to the same effective 4D theory derived from the 5D
  action which additionally depends on the particulars of the boundary
  conditions imposed. Different boundary conditions would typically
  induce distinct 4D theories as will be seen in a later section.}.
Specifically, the discrete theory defined in eq.\ref{eqn:lscal2} is
obtained with the assignments
\beq
\begin{split} \label{eqn:xydefn}
  & X_j \propto Z_j \propto \exp\left(\frac{-4k \pi R}{6p}|j-p|\right)\\[1ex]
  \mbox{and} \qquad & q \equiv \exp\left(\frac{k \pi R}{2p}\right), \qquad
  m^2 \equiv \frac{4p^2}{\pi^2 R^2 q} \ .
  \end{split}
\eeq
Therefore, in the continuum limit, $p\to \infty$, the 5D metric has the form,
\beq \label{eqn:cwmetr}
ds^2= 
\exp\left(\frac{-4}{3}k |z-z_p|\right) \, (dx^2+dz^2),
\eeq
where $z_p=\pi R/2$. 
Furthermore, writing the parameters $q$ and $m^2$ defined in eq.\ref{eqn:xydefn} as functions of the lattice spacing $a$, which acts as a natural UV regulator here, we have
\beq \label{eqn:paramrg}
q(a) = e^{k a} \quad \mbox{and} \quad
m^2(a) = \left[a^2 q(a)\right]^{-1}  \ .
\eeq
In the far UV ($a \to 0$) limit, an expansion in
  $a$, leads to a simple functional form for the mass eigenvalues of the
discrete theory, namely,
\beq
\begin{split} \label{eqn:discscalev}
m^2 \left[ 1+q^{2}-2q\cos\Big(\frac{n\pi}{p}\Big) \right] & \longrightarrow k^2 + \frac{4 n^2}{R^2} + \mathcal{O}(1/2p) \\
2m^2\left(1+q^2 \right) & \longrightarrow \infty \ .
\end{split}
\eeq
 In other words, these evolve into the KK
 spectrum of a scalar with an apparent bulk mass $k^2$. However, as can be easily ascertained, the scalar is actually massless, with this apparent ``bulk mass'' being an artefact of the warping.
Whether we obtain the exact same KK spectrum as above for the bulk
theory defined in eq.\ref{eqn:bscalaction} will be ascertained in a
subsequent section.

\subsection{The extended linear dilaton theory} \label{subsec:cwgeom}

Although the preceding treatment looks somewhat similar to that in
ref.\cite{Giudice:2016yja}, the precise outcomes of
eq.\ref{eqn:cwmetr} and eq.\ref{eqn:discscalev} specify a distinct
geometry and a consequent mass spectrum that is distinctively
different. However, until now we have just assumed the background
geometry and it is contingent on us to obtain the metric in
eq.\ref{eqn:cwmetr} as a solution of the five-dimensional Einstein
equations. To this end, we consider a 5D linear dilaton scenario
\cite{Antoniadis:2011qw,Cox:2012ee}, with the fifth dimension
compactified on a $S^1/\mathbb{Z}_2$ orbifold of radius $R$.
Furthermore, three rigid 3-branes are placed at $z=0, \, \pi R/2$ and
$\pi R$ respectively. The bulk and brane actions in this setup are, in
the Einstein frame,
\beq\label{eqn:ldgaction}
\begin{split}
  \mathcal{S}_{Bulk}&= 2 M_5^3
  \int d^4x \, dz \sqrt{-g} \left(\frac{1}{4}\mathcal{R}-\frac{1}{12} g^{MN} \partial_{M} S \partial_{N} S - V(S)\right) \\
  \mathcal{S}_{Brane}&= - 2M_5^3
  \int d^4x \, dz \frac{\sqrt{-g}}{\sqrt{g_{zz}}}\sum_{\alpha=1}^{3} \lambda_{\alpha}(S) \delta(z-z_\alpha) \ ,
\end{split}
\eeq
where $z_\alpha \equiv \{ 0,\, \pi R/2,\, \pi R\}$.
Here, $\mathcal{R}$ is the 5D Ricci scalar, $S(x,z)$ is the dilaton
field and $M_{5}$ is the 5D fundamental mass scale. The bulk and brane
potentials, $V(S)$ and $\lambda_{\alpha}(S)$ are given, respectively,
by
\beq
    V(S)=-e^{-2S/3 }k^2  \ , \qquad
    \lambda_{\alpha}(S)=\frac{e^{-S/3}}{2} \frac{\Lambda_{\alpha}}{M_{5}^{3}} \qquad ,
\eeq
where the $\Lambda$'s ($\Lambda_2=4 k
M_5^3=-\Lambda_1=-\Lambda_3$) are the vacuum energies on the branes.
The dimensionful parameter $k$ provides a measure of
  the vacuum energy in the bulk and, thereby, the extent 
  of the warping in the resulting geometry\footnote{A similar
5D setup was assumed in \cite{Kogan:1999wc, Agashe:2007jb} in the
context of RS-like warped scenarios, albeit without a linear dilaton.}.
The most general metric with 4D Poincare invariance corresponding to
the action in eq.\ref{eqn:ldgaction}, written in terms of the
conformal coordinates is~\cite{DeWolfe:1999cp},
\beq \label{eqn:ldgmet}
ds^2=e^{2 \sigma(z)}\left(\eta_{\mu \nu}dx^{\mu}dx^{\nu}+dz^2 \right)  , \qquad
\eta_{\mu \nu}=diag\{-1,+1,+1,+1 \}.
\eeq
The corresponding equations of motion
are given by
\beq\label{eqn:ldgeom}
\begin{split}
  \left(S''+3\sigma' S'\right)e^{-2\sigma(z)} &= 4k^2 e^{-S/3}- M_5^{-3} \sum_\alpha
  e^{-(\sigma+S/3)} \Lambda_{\alpha} \delta(z-z_\alpha) \\
\left[9\left(\sigma''-\sigma'^2 \right) + S'^2\right]e^{-2 \sigma(z)}&=-3M_5^{-3}\sum_\alpha e^{-(\sigma+S/3)} \Lambda_{\alpha} \delta(z-z_\alpha) \\
36 \sigma'^2-S'^2&=12 k^2 e^{2(\sigma-S/3)} \ .
\end{split}
\eeq
These are most easily solved by writing the bulk potential $V(S)$ in
terms of a  superpotential $W(S)$ \cite{DeWolfe:1999cp}, {\em viz.},
\beq
V(S)=\frac{3}{4}\left( \frac{\partial W(S)}{\partial S}\right)^2 -\frac{1}{3}W(S)^2 \, ,
\eeq
such that the solutions to the equations
\beq \label{eqn:suppot2}
    S'(z)=3e^{\sigma(z)}\frac{\partial W(S)}{\partial S} \ , \qquad 
    \sigma'(z)=-\frac{1}{3}e^{\sigma(z)}W(S)
\eeq
subject to the junction conditions 
\beq \label{eqn:ldgjcs}
W(S)\Bigg|^{z_\alpha+\epsilon}_{z_\alpha-\epsilon}=e^{-S/3}\frac{\Lambda_\alpha}{M_5^3}
\qquad \mbox{and} \qquad 
\frac{\partial W}{\partial S}\Bigg|^{z_\alpha+\epsilon}_{z_\alpha-\epsilon}=-\frac{1}{3}e^{-S/3}\frac{\Lambda_\alpha}{M_5^3} 
\eeq
(with $\epsilon$ being an infinitesimally small positive quantity)
are also solutions to eq.\ref{eqn:ldgeom}.
For our case, this implies
\begin{equation}
  W(S)=
  \left\{
  \barr{rcl}
\dis e^{-S/3}\sum_{\alpha=1}^{3}\frac{\Lambda_\alpha}{2 M_5^3}\left[\theta(z-z_\alpha)-\theta(z_\alpha-z)\right] &\qquad  &   z \in (0, \pi R), \\[3ex]
\dis e^{-S/3}\left[-\frac{\Lambda_1}{2 M_5^3}\theta(-z)+\frac{\Lambda_3}{2 M_5^3}\theta(z-\pi R) \right] & &  z \not\in (0, \pi R)\ .
\earr
\right.
  \label{eqn:superpot_form}
\eeq
With the orbifolding dictating that $\sigma(z)$ be symmetric about the
orbifold fixed points, the second of eqs.\ref{eqn:suppot2} implies that $W$ must be an odd
function across the fixed points. This, alongwith the conditions imposed by eq.\ref{eqn:ldgjcs},
fixes the form of $W(S)$ in the second line of
  eq.\ref{eqn:superpot_form}.  With these ingredients in order, we
obtain the following solutions to eq.\ref{eqn:ldgeom},
\beq \label{eqn:ldgsol1}
\begin{split}
  \sigma(z)&=-\frac{2}{3}k|z-z_p|e^{(\sigma_o-\frac{1}{3}S_o)}+\sigma_o \\
 S(z)&=-2k|z-z_p|e^{(\sigma_o-\frac{1}{3}S_o)}+S_o \ ,
 \end{split}
\eeq
where $\sigma_o$ and $S_o$ are constants of integration, and we use the notation $z_p=\pi R/2$.  Choosing, for
  convenience, $\sigma_o=S_o/3$, the solutions reduce
to\footnote{This can also be regarded as a rescaling of the parameter
  $k$ by defining
  $k=\tilde{k}e^{-(\sigma_o-\frac{1}{3}S_o)}$. Relating $k$ to the
  vacuum energy in the bulk demands that
  $\tilde{k}e^{-(\sigma_o-\frac{1}{3}S_o)}< M_5$. While $S_o$ is
  related to the VEVs of the dilaton at the branes as we will see in
  the next subsection, $\tilde{k}$ and $\sigma_o$ are free parameters
  which can be suitably configured to get optimum values for $k$.}
\beq \label{eqn:ldgsol2}
\begin{split}
  \sigma(z)&=-\frac{2}{3}k|z-z_p|+\sigma_o \\
 S(z)&=-2k|z-z_p|+S_o \ ,
 \end{split}
\eeq
The metric in eq.\ref{eqn:ldgmet}, therefore, has the explicit
form \footnote{Note that we drop the overall constant factor
  $e^{2\sigma_o}$ here as it does not affect the form of the
  geometry.}
\beq \label{eqn:ldgmetr2}
ds^2=\exp\left(\frac{-4}{3}k |z-z_p|\right) (dx^2+dz^2).
\eeq
So, the extended linear dilaton (LD) theory defined in
eq.\ref{eqn:ldgaction} successfully generates the required
\textit{clockwork} geometry of eq.\ref{eqn:cwmetr}. Evidently, the
symmetry of the triple brane setup about $z=z_p$ in terms of the
physical parameters in the theory has induced $\mathbb{Z}_2$ symmetric
solutions for $\sigma(z)$ and $S(z)$ and, hence, for the resulting
metric, under the transformation $z\to \pi R-z$.

\subsection{Addressing the hierarchy problem} \label{subsec:hier}
The role of the linear dilaton model as a solution to the hierarchy
problem has been elucidated in detail in
refs.\cite{Antoniadis:2011qw,Giudice:2016yja} and analogous arguments
readily follow for our three 3-brane setup. For brevity's sake, we
refrain from discussing this in detail and, instead, highlight only
the key differences. To begin with, starting with the fundamental
scale $M_5$, we may obtain, on dimensional reduction, the Planck mass
$(M_P \sim 10^{19})$ as the effective mass scale in the 4D
Einstein-Hilbert action, {\em viz.},
\begin{equation}  \label{eqn:ldgeff}
M_{P}^2 =2M_5^{3}\int_{0}^{\pi R} dz \, e^{3\sigma(z)} 
= \frac{2 M_5^{3}}{k}\left(1-e^{-k \pi R} \right).
\end{equation}
The above relation resembles the one obtained in the RS
model~\cite{Randall:1999ee}.

The major difference from the two-brane
theory lies in the emergence of two distinct physical scenerios that
ameliorate the hierarchy issue. For $k>0$, we have two negative tension (IR)
branes at the orbifold boundaries and a positive tension (UV) brane at
the central $\mathbb{Z}_2$ fixed point. We shall call it the IR-UV-IR
setup.  In this scenario, we have $M_P \sim M_5$ for a nominally large
value of $k \pi R \sim \mathcal{O}(10)$. For a given value of $k$, though, the required magnitude of $k\pi R$
is roughly twice of that necessary in two-brane models. This difference stems from the fact that in
our three brane model the hierarchy is explained by
the warping profile operative between the boundary branes and the
brane present in the middle of the orbifold. A standard model (SM) sector localised on the IR brane would have its
effective mass parameter\footnote{For the SM this means the
  independent mass parameter $\mu$ in the Higgs sector or,
  equivalently, the symmetry breaking scale $v$.} ($m_{IR}$) derived
from the parameters in the fundamental 5D theory via relations like
$m_{IR}=e^{-a k \pi R} m_{0}$, thereby lowering it to the TeV
scale. Here, $a$ is an $\mathcal{O}(1)$ numerical constant arising from a combination of the exponential in the (induced) 4D metric determinant and that emanating from field redefinitions. $m_0$
denotes a parameter of the brane localised sector in the full 5D
theory, presumably near the Planck scale. Note that this argument is
similar to the one presented in the original RS scenario
\cite{Randall:1999ee}.

On the other hand, for $k<0$, the negative tension brane is located at
the central fixed point and the positive tension branes at the
orbifold boundaries. Hence, we refer to this as the UV-IR-UV
setup. The hierarchy is addressed in a similar manner as before for
$|k \pi R|\sim \mathcal{O}(10)$, only that in this case the fundamental scale, $M_5$, is near
the TeV scale.

The triple brane theory, therefore, proposes the plausibility of two
distinct scenarios in terms of their role in explaining the Planck-EW
hierarchy (Fig.\ref{fig:ldg}). However, one needs to check for the
stability of both the scenarios in terms of the physicality of the
emerging dilaton modes. This is necessary as the coefficients of their
kinetic terms depend on the parameters of the theory and, hence, may
lead to unphysical degrees of freedom for certain values of the
parameter space. For a related discussion see Appendix
\ref{sec:dyndil}.

Phenomenologically, the two scenarios could have distinct, albeit
rich, consequences. For instance, the dynamical dilaton (see Appendix
\ref{sec:dyndil}) and graviton modes emerging in the two setups would
lead to their different couplings with the SM confined to the IR
brane. A more
  detailed inspection of these aspects, though, is beyond the scope of
  the current exercise and would be pursued later.

\begin{figure*}[tbp] 
\centering
      \includegraphics[scale=0.9,keepaspectratio=true]{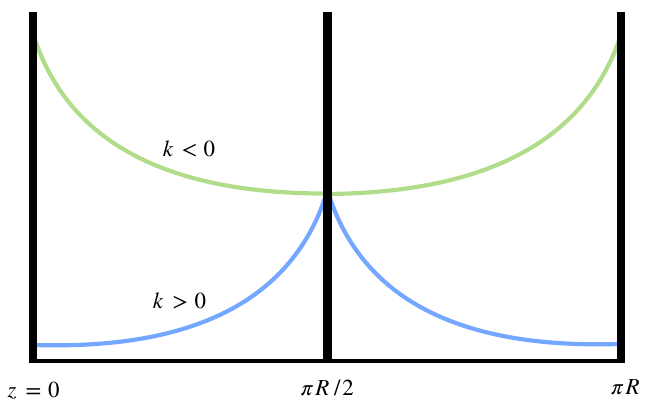}
	\caption{Graphical depiction of the warp factor $e^{2\sigma(z)}$ in the IR-UV-IR (blue) and the UV-IR-UV (green) scenario.} \label{fig:ldg}  
\end{figure*}

\subsection{Radius stabilisation}

Modulus stabilisation in warped scenarios like the RS model typically
involves the addition of a bulk scalar field (the modulus field) whose
VEVs at the branes stabilise the size of the extra dimension \textit{$\grave{a}$ la} the Goldberger-Wise mechanism \cite{Goldberger:1999uk}. In a
linear dilaton model the dilaton field itself can accomplish the role
of a modulus field\footnote{Even if the dilaton field is not
  considered to be responsible for the stabilisation, a GW-like
  mechanism can be consistently adopted for the purpose. See
  \cite{Giudice:2016yja} for a discussion.} \cite{Cox:2012ee}. In this
section we extend those arguments to our triple brane
model\footnote{Stabilization of triple-brane systems have been studied
  before in \cite{Choudhury:2000wc} in the context of the RS
  scenario.} by invoking additional potentials on the branes. These potentials could, in principle, arise even as quantum
corrections on the brane resulting from the dilaton's interactions
with brane localised matter. Rather than posit a complicated
potential, we limit ourselves to the simplest choice, {\em viz.}
\begin{equation} \label{eq:brane_pt}
\lambda_{\alpha}(S)=\frac{e^{-S/3}}{2} \left[\frac{\Lambda_{\alpha}}{M_{5}^{3}} + \mu_{\alpha}\left( S-\mathcal{V}_\alpha\right)^2 \right],
\end{equation}  
where $\mu_{\alpha}$ and $\mathcal{V}_\alpha$ are constants. For this to be consistent with the junction conditions
in eq.\ref{eqn:ldgjcs} we must have
\begin{equation} \label{eq:radjcs}
S(z_\alpha)=\mathcal{V}_\alpha.
\end{equation} 
In other words, the junction conditions impose boundary conditions on
the dilaton profile such that the dilaton field acquires nonzero
vacuum expectation values at the branes. This further ensures that the
augmented brane potentials do not generate any backreaction in the
system.

Now, taking the general form of the solutions in
eq.\ref{eqn:ldgsol2} and imposing the appropriate boundary
conditions at the end-of-the-world branes, {\em viz.},
\begin{equation}
S(z_{\alpha})=-k\pi R + S_o=\mathcal{V}_{\alpha} \ ,
\end{equation}
(where $S_o = \mathcal{V}_2$ from eq.\ref{eq:radjcs})
we obtain the size of the extra dimension in terms of the vacuum
expectation values of the dilaton field at the branes
\begin{equation} \label{eqn:ldgrad1}
\pi R = \frac{\left( \mathcal{V}_{2}-\mathcal{V}_{1} \right)}{k}
\qquad \mbox{and} \qquad
\pi R = \frac{\left( \mathcal{V}_{2}-\mathcal{V}_{3} \right)}{k} \ .
\end{equation}
However, note that, together, the two conditions stipulate that the
mechanism is consistent only when the VEVs on the branes located at
the orbifold boundaries are equal. While this can be ensured by
imposing an exact $\mathbb{Z}_2$ symmetry (about $z=z_p$) in the full
(bulk--brane) theory, it would be interesting to find a mechanism or a
more fundamental theory in which such a symmetry would emerge
naturally. In the purview of our current discussion, though, the
imposition $\mathcal{V}_1=\mathcal{V}_3$ might (naively) suggest an
additional fine-tuning in the theory, an aspect we comment upon in
section \ref{sec:ft}.

\section{Bulk fields in clockwork/linear dilaton geometry}
\label{sec:contcw}
Having determined the configuration of the 5D theory which generates
the modified \textit{clockwork} geometry, it is imperative to study
the nature of bulk field theories in the background specified by the
metric in eq.\ref{eqn:ldgmetr2}, so as to ascertain and verify the
correspondence drawn earlier between the discrete and the 5D continuum
theories. Although it is straightforward to generalise the ensuing
discussion to non-Abelian theories we choose to study only bulk
Abelian theories here in order to directly compare with the discrete
clockwork theories. As before, we limit the discussion to only bosonic
fields. We note that, in eq.\ref{eqn:ldgaction}, the terms
proportional to $M_5^{3} k^2$ are the ones that primarily influence
the geometry. In the following bulk field actions neither do we assume
any large mass parameters nor do we assume vacuum expectation values
for the scalar fields.  Consequently, the backreaction due to these
fields can be entirely neglected.

\subsection{Bulk scalar in LDG}
Assuming the metric of eq.\ref{eqn:ldgmet}---a valid assumption
as long as the backreaction due to the
new field is negligible---the
  action for a scalar field $\phi$ is given by
\beq \label{eqn:contscal1}
\mathcal{S}= \frac{-1}{2}\int d^4x \, \int_0^{\pi R}dz \, e^{3\sigma(z)}
        \left[(\partial_\mu \phi)^2 + (\partial_z \phi)^2\right]
\eeq
which is invariant under the $\mathbb{Z}_2$ transformation ($z \to \pi
R -z$) with $z = z_p$ being a fixed point under this
transformation\footnote{This, of course, is in addition to the
  orbifold fixed points.}. The existence of this symmetry stipulates
that the solutions of the equation of motion (EOM) should be
classifiable as representations of the corresponding parity operator
$\mathcal{Z}$, namely\footnote{Note that we treat the fields
  $\phi^{(\pm)}(x,z)$ as two independent degrees of freedom which
  enables us to simultaneously impose different sets of boundary
  conditions for the even and odd fields.  This follows from the fact
  that, in the two patches of the orbifold, the fields can be deemed
  to be {\em a priori} disconnected (and, hence, independent) with
  support only on the individual patches. Demanding continuous
  solutions to the EOM, the
  $\mathbb{Z}_2$ symmetry of the system dictates the existence of two
  (even and odd) fields $\phi^{(\pm)}(x,z)$ as linear combinations of
  the piecewise fields. Hereinafter, we resort to discussing the
  physical properties of the theory directly in terms of the even and
  odd bulk fields.}
\beq \label{eqn:bscalpar}
\mathcal{Z} \phi^{(\pm)}(x, z ) \equiv  \phi^{(\pm)}(x, \pi R-z )
= \pm \phi^{(\pm)}(x, z ) \ .
\eeq
Clearly, this is analogous to the $\mathbb{Z}_2$ parity in the
discrete case. The $U(1)_{CW}$ symmetry in the discrete theory---see
eq.\ref{eqn:cwgen2}--- is reflected by the invariance of the action
under a constant shift of the bulk fields, $\phi^{(\pm)}(x,z) \to
\phi^{(\pm)}(x,z) + c$.  Effecting the KK decompositions
\beq
\phi^{(\pm)}(x,z)= \frac{1}{\sqrt{\pi R}} \,
      \sum_{n=0}^\infty \, \varphi_n^{(\pm)}(x) f^{(\pm)}_n(z) \ ,
\eeq
with the obvious normalisation
\beq
\frac{1}{\pi R}\int_0^{\pi R}dz \, e^{3 \sigma(z)}f^{(\pm)}{(n)} f^{(\pm)}{(n')} = \delta^{n n'},
\eeq
the EOMs for
$\varphi^{(\pm)}(x)$ and $f^{(\pm)}(z)$ are
\begin{equation}  
  \barr{rcl}
  \dis \left(\partial_x^2 - m_n^2 \right) \varphi^{(\pm)}_n(x) & = & 0 \\[2ex]
  \dis \left(\partial_z^2 -k^2 + m_n^2 \right)
    \left(e^{3 \sigma(z)/2}f^{(\pm)}_n(z)\right) & = & 0 \ .
  \earr
  \label{eqn:bscaleom}
\end{equation} 
At this stage, the masses $m_n$ for the KK modes $\phi_n$ are
arbitrary, and are to be determined only by solving for the
eigenvalues of the operator in the second equation. This, in turn,
needs the boundary conditions (BC) on the bulk field to be specified.
With the $\mathbb{Z}_2$ stipulating that the BC at $z=2z_p$ must be
given by that at the origin, we choose to impose the BCs at $z = 0$
and $z=z_p$. Given that an individual BC could either be
Dirichlet-like ($\phi(x,z)=0$) or Neumann-like ($\partial_z
\phi(x,z)=0$), various combinations are possible, each with its
particular set of consequences, and we examine below the two broad
classes:
\begin{itemize}
\item \textbf{Case I: } Imposing {\em identical} BCs at $z=0,z_p$, we
  have two further choices, namely both Dirichlet-like (DD) or both
  Neumann-like (NN) Explicitly, we obtain the following solutions:
\\
\underline{\textbf{Neumann BC at both $z=0 \, \mbox{and} \, z_p$ (NN)}} \\
This admits only the even solutions, namely (here, $n \geq 1$)
\beq \label{eqn:scaleven}
\barr{rcl}
\dis f_0^{+}(z)&=& \dis \mathcal{N}_0 \\
m_0&=& 0\\
f_n^+(z) &=& \dis
\mathcal{N}_{n}^{+} e^{-\frac{3}{2}\sigma(z)}\left\{-\frac{kR}{2n} \sin\left[\frac{2n}{R} |z-z_p|\right]+\cos\left[\frac{2n}{R} (z-z_p)\right]\right \} \\
m_n^2 &=& \dis k^2 + \frac{4 n^2}{R^2}   \ .
\earr 
\eeq

\underline{\textbf{Dirichlet BC at $z=0 \, \mbox{and} \, z_p$ (DD)}} \\
Understandably, these lead only to the odd solutions, namely 
\beq \label{eqn:scalodd}
f_n^-(z)= \mathcal{N}_{n}^{-} e^{-\frac{3}{2}\sigma(z)}\sin\left[\frac{2n}{R} (z-z_p)\right] \ , \qquad 
m_n^2 = k^2 + \frac{4 n^2}{R^2} \ .
\eeq
The normalisation factors are given by
\begin{equation}
\mathcal{N}_0 =\sqrt{\frac{k \pi R}{1-e^{-k \pi R}}}, \qquad 
\mathcal{N}^{+}_{n}= 2\sqrt{2} \frac{n}{m_n R}, \qquad \mathcal{N}^{-}_{n}= \sqrt{2} 
\end{equation}
 
Thus, the physical spectrum consists of an exactly massless mode
($\varphi_0$) and, beyond a mass gap of magnitude $\sim k$, two
mass-degenerate KK-towers, with one each of $\mathbb{Z}_2$-even
($\varphi_n^{(+)}$) and odd ($\varphi_n^{(-)}$) mode at each level. In
a limited sense, this mimics the spectrum of the discrete theory (see
eq.\ref{eqn:discscalev}). Just as in the discrete case, the
occurrence of a mass degeneracy can be attributed to the
$\mathbb{Z}_2$ symmetry of the action as well as to the boundary
conditions assumed. (For further elucidation of the mass degeneracy,
see Appendix \ref{subsec:condeg}.) Note that, given the BCs, the
EOM has no non-trivial solution for $m^2 < k^2$.

The mass eigenvalues are graphically shown in Fig.\ref{fig:spec_cont}
for a benchmark set of values for the parameters $k$ and
$R$. Evidently, the inter-level spacing increases progressively with
$n$ and then tends to saturate to $\Delta m_n \sim 2\pi/R$ for large
$n$. While this might seem at variance with the discrete case, this
was only to be expected. If the gravitational interaction could be
neglected, the five-dimensional theory could, notionally, be treated
as a free and, hence, UV-complete one with the masses defined by $k$
and $R^{-1}$. The spectrum of the discrete theory, on the other hand,
must depend on both the lattice spacing $a$ and the UV-scale $f$.
Yet, in the event of $p$ being large, an analogous behaviour is seen
for $n \ll p$ with an opposite behaviour (namely, the subsequent
spacings decreasing) being seen for $n > p$ (see
eq.\ref{eqn:discscalev}). In other words, the mass eigenvalues therein
assume the form of the KK spectrum when the RG evolution of the
parameters, with respect to the lattice spacing $a$, is taken into
account in the sense indicated in eq.\ref{eqn:discscalev}.

Now that we have obtained the solutions, the theory in
eq.\ref{eqn:contscal1} can be rewritten in terms of the even and odd
bulk fields. However, this would tend to make the action look
cumbersome as one adds higher order terms.  Since, all that we expect
after dimensional reduction is a 4D theory with the $\mathbb{Z}_2$
parity intact, it is simpler to consider the equivalent, namely a
single bulk field $\phi(x,z)$ and ascribe it with the expansion
\beq
\phi(x,z)\equiv \frac{1}{\sqrt{\pi R}}\sum_{n=0}^\infty \left\{\varphi_n^{(+)}(x) f_n^{(+)}(z)+\varphi_n^{(-)}(x) f_n^{(-)}(z) \right\}
\eeq
when the 4D theory is to be determined.

With this definition, we examine the coupling profile of the bulk
sector with an operator $\mathcal{O}_E$ confined to {\em one} of the branes,
{\em i.e.}, at $z= z_\alpha \equiv \{0, \pi R/2, \pi R \}$. To this end, we introduce the interaction action 
\begin{equation} \label{eqn:coupext1}
\begin{split}
  \mathcal{S}_{ext}=-\int d^4x \, dz \, \frac{\sqrt{-g}}{\sqrt{g_{zz}}}
  \exp\left(\frac{-5}{6}S(z)\right) \mathcal{O}_E \phi(x,z)\delta(z-z_\alpha)
\end{split}.
\end{equation}
The dilaton factor $e^{- 5\, S(z)/6}$ has been introduced to induce
localisation of the $0$-th mode towards one of the fixed points. The
numerical constant ($-5/6$) is fixed by demanding that the heavy KK
modes are de-localised, i.e., the exponential factor in their
wavefunctions is eliminated. This choice, although not strictly
necessary in the context of the 5D theory alone, is adopted to mimic
the coupling profile typical of the clockwork mechanism in the
corresponding discrete theory\footnote{In general we can also
  introduce a dilaton factor $e^{cS(z)}$ coupling to the overall
  Lagrangian in the bulk action of eq.\ref{eqn:contscal1}, where $c$
  is a constant which determines the coupling strength. Such a
  construction has been studied in \cite{Kang:2020cxo} in the context
  of the original CW/LD geometry. We assume $c=0$ in our modified
  theory only to simplify matters in hand and generalisation with
  respect to dilaton couplings would be an interesting aspect to
  explore.}. Using eq.\ref{eqn:scaleven}-\ref{eqn:scalodd}, the above
action simplifies to
\beq
\barr{rcl}
\dis \mathcal{S}_{ext}&=& \dis\frac{-1}{\sqrt{\pi R}}\int d^4x \, dz \, \frac{\sqrt{-g}}{\sqrt{g_{zz}}} \, \exp\left(\frac{-5}{6}S(z)\right) \delta(z-z_\alpha)
\\[1.5ex]
& & \dis \hspace*{6em}
\left\{\sum_{n=0}^{\infty}f^{+(n)}\mathcal{O}_E \varphi^{+(n)}(x) + \sum_{n=0}^{\infty}f^{-(n)}\mathcal{O}_E \varphi^{-(n)}(x)\right\} \\[3ex]
&=& \dis \frac{-1}{\sqrt{\pi R}}\int d^4x \Bigg\{\mathcal{N}_0 e^{-k|z_\alpha-z_p|}\mathcal{O}_E \varphi^{+(0)}(x) \\
& &\dis \hspace*{4em} + \sum_{n=1}^{\infty}\mathcal{N}^{+}_{(n)} \mathcal{O}_E \varphi^{+(n)}(x) \\[2ex]
& & \dis \hspace*{6em} \left[ -\frac{kR}{2n}\sin\left(\frac{2n}{R}|z_\alpha-z_p|\right) 
  +\cos\left(\frac{2n}{R}(z_\alpha-z_p)\right)\right]\Bigg\}.
\earr
\eeq
The fact that the odd wavefunctions vanish at the fixed points and
that the external operator is trivially invariant under the bulk
$\mathbb{Z}_2$ tranformation ensure that the brane localised
interactions preserve the $\mathbb{Z}_2$ parity. Evidently, the
coupling of the zero mode at the boundary branes ($z_\alpha=0, \pi R$)
is exponentially suppressed, whereas, that at the central brane
($z_\alpha=z_p$) is $\sim \mathcal{O}(1)$. In contrast, the massive
modes have nearly similar couplings, of magnitude $\sim
\mathcal{O}(1)$, at every brane.  This, clearly, is analogous to what
we obtain in the discrete case discussed in
sec.\ref{subsec:discscal}.

\item \textbf{Case II:} An interesting possibility is afforded by the
  imposition of an \textit{asymmetric} set of BCs, e.g., Neumann BC at
  $z=0$ and Dirichlet BC at $z=z_p$ or vice-versa. These BCs do not
  admit an exactly massless mode which is in stark contrast with what
  we encounter in case \textbf{I}.  Explicitly, 

  \underline{\textbf{Neumann BC at $z=0$ and Dirichlet BC at $z=z_p$
      (ND)}} leads to the odd solutions
  
\beq \label{eqn:scalodd2}
f_n^-(z) = \tilde{\mathcal{N}}_{n}^{(-)} e^{-\frac{3}{2}\sigma(z)}\sin\left[\beta_n (z-z_p)\right]
\eeq
where
$\beta_n$ are 
the solutions to the trascendental equation
\beq
\tan{\frac{\beta_n \pi R}{2}} = -\frac{\beta_n}{k}.
\eeq
and $m_n^2 = k^2 + \beta_n^2$. On the other hand,

\underline{\textbf{Dirichlet BC at $z=0$ and Neumann BC at $z=z_p$ (DN)}}, gives the even solutions
\beq \label{eqn:scaleven2}
f_n^+(z)=\tilde{\mathcal{N}}_{n}^{(+)} e^{-\frac{3}{2}\sigma(z)}\left\{-\frac{k}{\beta_n} \sin\left[\beta_n |z-z_p|\right]+\cos\left[\beta_n (z-z_p)\right]\right \}
\eeq
Once again, $\beta_n$ are given by a very similar equation, namely,
\beq
\tan{\frac{\beta_n \pi R}{2}} = \frac{\beta_n}{k}.
\eeq
with $m_n^2 = k^2 + \beta_n^2$.  That the (ND) and (DN) cases admit
odd and even solutions, respectively, is dictated by the boundary
conditions and the continuity conditions at $z=z_p$ in each case,
along with the fact that the theory is $\mathbb{Z}_2$
symmetric. Clearly, the KK modes are non-degenerate with the even
modes being heavier than the odd ones for a particular KK level, and,
hence, incommensurable with the discrete CW mass spectrum in the sense
shown\footnote{It remains to be examined whether this scenario
  has a different (non-CW) discrete analogy that could be derived as a
  deconstruction of the 5D theory, presumably by invoking a
    different discrete form for the $z$-derivatives than what we have
  adopted here.} in eq.\ref{eqn:discscalev}. Similar to case
\textbf{I}, the inter-level spacings in the individual towers increase
as we go higher in the spectrum and tend to saturate for large values
of $n$ for which $\beta_n \to (2n-1)/R \gg k$. Therefore, any
quasi-degeneracy seemingly present in the low lying KK states gets
disturbed in the heavier levels.

In addition to the KK spectrum,  there exists a single non-trivial
$\mathbb{Z}_2$ even solution (satisfying the \textbf{DN} BCs) for $m^2
< k^2$ given by
\beq \label{eqn:scaleven3}
\begin{split}
f_0^+(z)&=\tilde{\mathcal{N}}_{0}^{(+)} e^{-\frac{3}{2}\sigma(z)}\left\{-\frac{k}{\tilde{\beta}} \sinh\left[\tilde{\beta} |z-z_p|\right]+\cosh\left[\tilde{\beta} (z-z_p)\right]\right \}
\end{split}.
\eeq
where $\tilde{\beta}\equiv \sqrt{k^2-m_0^2}$ is obtained as a solution of the equation
\beq
\tan{\frac{ \tilde{\beta} \pi R}{2}} - \frac{\tilde{\beta}}{k}=0.
\eeq

This is unlike case \textbf{I} which admits a trivial massless mode.
For values of $|k \pi R|$ that potentially address the EW-Planck
hierarchy, the preceding equation has the solution $\tilde{\beta} \sim
k$, with $f_0^+(z) \to \tilde{\mathcal{N}}_{0}^{(+)}$, {\em i.e.}, an
{\em almost} flat profile along the fifth dimension similar to what
was obtained for the exactly massless state in case \textbf{I}. In
other words, the erstwhile flat (and, hence, massless) solution
acquires a slight non-triviality in its wavefunction and, thereby, is
very slightly lifted in mass. A part of the full spectrum is shown
schematically in Fig.\ref{fig:spec_cont} for a specific set of values
for $k$ and $R$.

We now turn to the question of the coupling with a brane localised
operator. If an interaction of the form shown in eq.\ref{eqn:coupext1}
is assumed, the nature of coupling with the light mode remains quite
similar to what we encountered in case \textbf{I} for $|k \pi R| \gg
1$. The heavier KK modes, too, have similar ($\mathcal{O}(1)$)
couplings at the branes as previously seen. The major distinction,
however, appears in the case of the odd modes in that they now have
vanishing couplings only at the central fixed point. Couplings with a
sector confined to one of the boundary branes, therefore, would break
the $\mathbb{Z}_2$ parity explicitly unless there exist couplings with
a mirror sector in the opposite brane as well.

\end{itemize}
\begin{figure*}[tbp] 
\centering
      \includegraphics[scale=0.4,keepaspectratio=true]{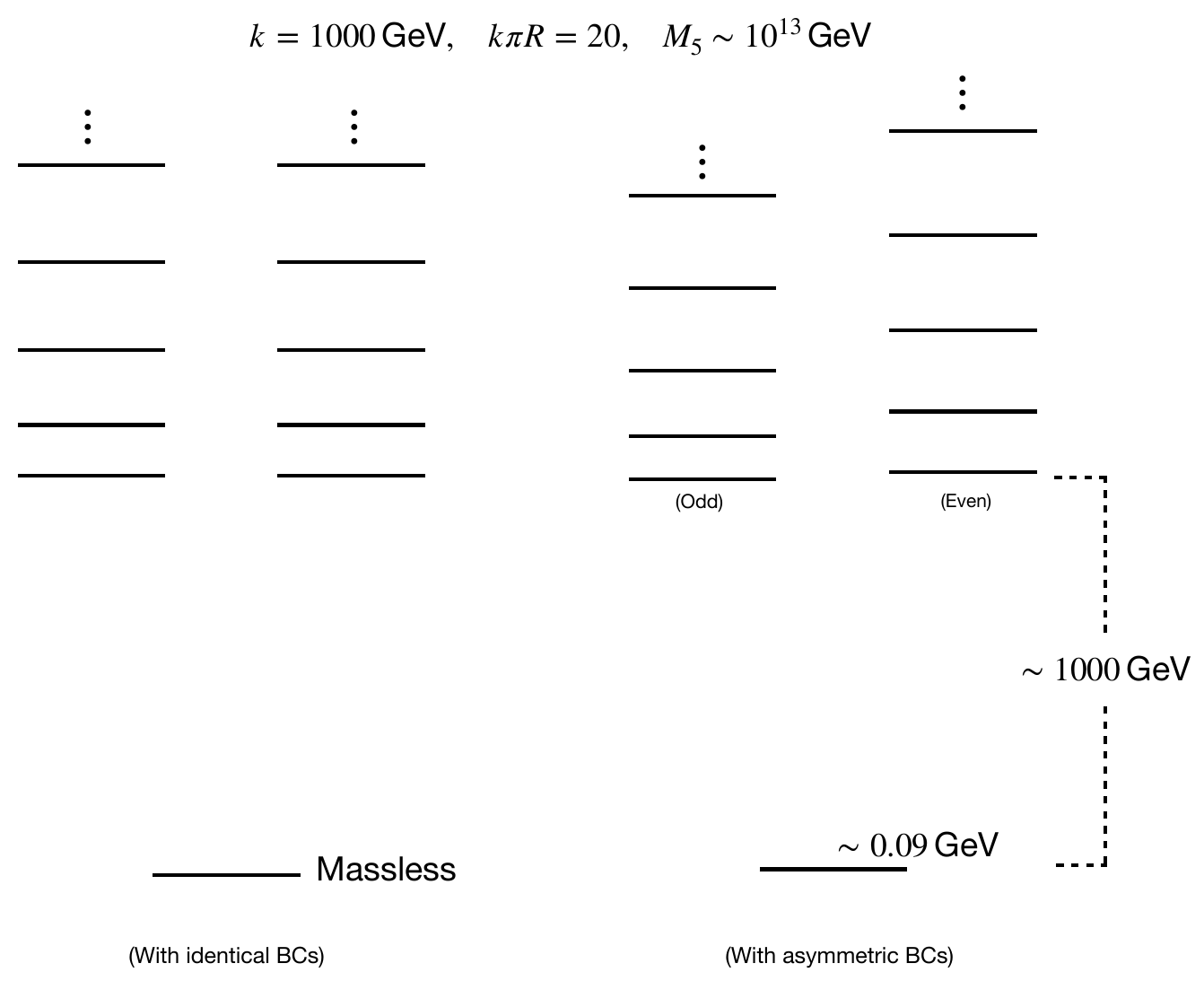}
	\caption{Schematic representation of the first few KK levels
          in the case of identical (left) and asymmetric (right)
          boundary conditions for a benchmark parameter
          set.} \label{fig:spec_cont}
\end{figure*}

At this juncture, an interesting distinction between the two sets of
BCs is worth mentioning. Note that the \textbf{(NN)} BC preserves the
symmetry under a constant shift of the 4D field $\varphi_0$, i.e.,
under $\varphi_0(x)+c$, due to the $z$-independence of its
wavefunction. This, in turn, justifies its masslessness. The
\textbf{(DD)} BC clearly breaks this symmetry which is why we do not
obtain an odd massless mode. The \textbf{(DN)} BC, too, breaks this
symmetry for the zeroth mode, thereby, leading to it being massive.

\subsection{Bulk vector in LDG}
The discussion of the bulk theory for a massless vector boson runs, to
a large extent, parallel to that for the scalar theory. The free field
action is, of course, given by
\begin{equation}
\mathcal{S}=\int d^4x \, dz \sqrt{-g}\left\{ -\frac{1}{4}g^{M M'}g^{N N'}F_{M N}F_{M' N'}   \right\}
\end{equation}
where $M = \{\mu, z\}$ {\em etc.}  Choosing the generalized $R_\xi$
gauge-fixing term
\begin{equation}
\mathcal{S}_{GF}=\int d^4x \, dz \sqrt{-g}e^{-4\sigma(z)}\left\{ -\frac{1}{2\xi}\left[\partial^\mu A_\mu + \xi e^{-\sigma(z)}\partial_z\left(e^{\sigma(z)}A_z\right)\right]^2\right\}
\end{equation}
allows us to eliminate all the $A_\mu-A_z$ mixing terms upto total derivatives along the $z$ dimension. We define the field decomposition to be
\begin{equation}
    A^{(\pm)}_\mu = \frac{1}{\sqrt{\pi R}}\sum_{n=0}^{\infty}\mathcal{A}^{(\pm)}_{\mu \, n}(x)f^{(\pm)}_{n}(z) \ , \qquad \qquad 
   A^{(\pm)}_z  = \frac{1}{\sqrt{\pi R}}\sum_{n=0}^{\infty}\mathcal{A}^{(\pm)}_{z \, n}(x)h^{(\pm)}_{n}(z)
\end{equation}
with the normalization conditions being
\begin{equation}
  \frac{1}{\pi R}\int dz\, e^{\sigma(z)}f^{(\pm)}_{n}f^{(\pm)}_{n'} =\delta_{n n'} \, ,
\end{equation}
where the choice $h_n \equiv (1/m_n)\partial_z f_n$ for $n>0$ is
dictated by the $\mathbb{Z}_2$ symmetry in the $A_\mu-A_z$ mixing terms
and $h_0 \equiv f_0$ is motivated by the fact that the zero mode
solutions of the respective EOMs are identical. Dropping
  the $(\pm)$ superscripts for brevity, we obtain the following EOMs for
  $\mathcal{A}_{\mu}^{(n)}(x)$, $\mathcal{A}^{(n)}_{z}$ and
  $f^{(n)}(z)$ (for both even and odd modes),
\begin{equation} \label{eqn:eom_vector}
\begin{split} 
\partial_\mu \mathcal{F}^{\mu \nu (n)} + m_{n}^2 \mathcal{A}^{\nu (n)} + \frac{1}{\xi}\partial^{\nu}\left(\partial^\mu \mathcal{A}_{\mu}^{(n)}\right) &=0 \\[1ex]
\left( \partial_{x}^{2} - \xi m_{n}^2 \right)\mathcal{A}_z^{(n)}(x)&=0 \\[1ex]
\left( \partial_{z}^{2} - \frac{1}{9}k^2 + m_{n}^2 \right)\left( e^{\sigma(z)/2} f^{(n)}(z)\right) &=0 \ ,
\end{split}
\end{equation}
where the field strength tensors $\mathcal{F}^{(n)}$ correspond to the
4D fields $\mathcal{A}_{\mu}^{(n)}$.  The equation for $f(z)$ is
similar to the one for the bulk scalar with the only difference being
in the coefficient of the $k^2$ term. Following the terminology
therein, the solutions for the \textit{identical} BC case are:

\noindent
\underline{\textbf{NN} boundary conditions}
\begin{equation} \label{eqn:veceve}
\begin{split}
f^{+ (0)}(z) &= \mathcal{N}_{0}^{(+)} \\
f^{+ (n)}(z) &=\mathcal{N}_{n}^{(+)} e^{-\sigma(z)/2}\left\{-\frac{k R}{6 n} \sin\left[\frac{2n}{R} |z-z_p|\right] + \cos\left[\frac{2n}{R} (z-z_p)\right]\right \} \\
m_0^2 &= 0\\
m_n^2 &= \frac{1}{9} k^2 + \frac{4 n^2}{R^2}
\end{split}
\end{equation}
\noindent
\underline{\textbf{DD} boundary conditions}
\begin{equation}   \label{eqn:vecodd}
f^{- (n)}(z) = \mathcal{N}_{n}^{(-)} e^{-\sigma(z)/2}\sin\left[\frac{2n}{R} (z-z_p)\right] \ , \qquad
m_n^2 = \frac{1}{9} k^2 + \frac{4 n^2}{R^2} \ ,
\end{equation}
where, as before, $n \geq 1$ and $(\pm)$ denote $\mathbb{Z}_2$
even and odd wavefunctions respectively. The normalisation
factors, $\mathcal{N}$, are given by
\begin{equation}
\begin{split}
\mathcal{N}_0=\sqrt{\frac{ k \pi R}{3\left(1-e^{-\frac{1}{3}k \pi R} \right)}}, \qquad \mathcal{N}^{+}_{n}=2\sqrt{2} \frac{n}{m_n R}, \qquad \mathcal{N}^{-}_{n}= \sqrt{2}
\end{split}.
\end{equation}
As in the scalar case, and as expected, we have exactly one massless
mode $A^{(+)}_{0\mu}$ which is to be identified with the
four-dimensional gauge field.  The 4D masses of the vector bosons can
be attributed to the Goldstone-like degrees of freedom
$\mathcal{A}_z^{(n)}$ $(n \geq 1)$ which, in the unitary gauge ($\xi \to \infty$), vanish from the effective 4D theory.  In
addition, there remains a massless scalar mode in the
spectrum. Therefore, following the treatment in the scalar theory, the
spectrum of even and odd KK modes can be classified in terms of the
$\mathbb{Z}_2$ parity,

\begin{equation}
  \mathcal{Z} \mathcal{A}_{\mu}^{(\pm) (n)}(x) = \pm \mathcal{A}_{\mu}^{(\pm) (n)}(x)
      \ , \qquad 
\mathcal{Z} \mathcal{A}_z^{(+) (0)}(x) = \mathcal{A}_z^{(+)  (0)}(x) \ .
\end{equation}

In other words, the $\mathbb{Z}_2$ assignments as well as the
degeneracy within a level is exactly the same as that for the spectrum
obtained in the discrete scenario in sec.\ref{subsec:disvec}. The
coupling profile of the KK modes with a brane localised operator can
be seen by assuming that the operator is a current ($\mathcal{J}_M=\{
\mathcal{J}_\mu, \mathcal{O}_{S}\}$) composed of fields charged under
the bulk $U(1)$ gauge group and, therefore,
couples to the bulk gauge field. The appropriate interaction is
defined by
\begin{equation} \label{eqn:bvecex}
\begin{split}
\mathcal{S}_{ext}&=-\int d^4x \, dz \, \frac{\sqrt{-g}}{\sqrt{g_{zz}}}e^{-\frac{7}{6}S(z)}\mathcal{J}_M A^{M}(x,z) \delta(z-z_\alpha),
\end{split}
\end{equation}
where $\mathcal{J}_\mu$ is a vector current \footnote{Identifying the
  external sector with the SM, this can be envisaged as the current
  associated with some of the popular and well-motivated scenarios
  with an Abelian gauge group like $U(1)_{B-L}$, $U(1)_{L_{\mu} -
    L_{\tau}}$, etc., or with a dark sector group $U(1)_D$ with the
  light vector being a $Z'$ or a dark photon-like particle. The light
  scalar, on the other hand, can be identified as an axion-like
  particle. Prospective models based on such bulk scenarios,
  therefore, would simultaneously induce couplings of a $Z'$ (or $A'$)
  and an ALP with the SM sector.}, $\mathcal{O}_S$ is a scalar
operator, and, similar to the scalar theory, we assume
  \beq
  \begin{split}
   A_{\mu}(x,z) & \equiv \frac{1}{\sqrt{\pi R}}\sum_{n=0}^\infty
  \left\{\mathcal{A}^{(+)}_{\mu \, n}(x)
  f_n^{(+)}(z)+\mathcal{A}^{(-)}_{\mu \, n}(x) f_n^{(-)}(z)\right\}   \\
  A_{z}(x,z) & \equiv \frac{1}{\sqrt{\pi R}}\sum_{n=0}^\infty
  \left\{\mathcal{A}^{(+)}_{z \, n}(x)
  h_n^{(+)}(z)+\mathcal{A}^{(-)}_{z \, n}(x) h_n^{(-)}(z)\right\} 
  \end{split}.
  \eeq
With this expansion, the action simplifies to (in the
  unitary gauge)
\begin{equation} 
\begin{split}
\mathcal{S}_{ext}
&=-\frac{1}{\sqrt{\pi R}}\int d^4x \, dz \, \frac{\sqrt{-g}}{\sqrt{g_{zz}}}e^{-\frac{7}{6}S(z)}\Bigg\{\sum_{n=0}^{\infty}f^{+(n)}\mathcal{J}_\mu \mathcal{A}^{\mu +}_{(n)}(x) + \sum_{n=0}^{\infty}f^{-(n)}\mathcal{J}_\mu \mathcal{A}^{\mu -}_{(n)}(x) \\ 
&+ f^{+(0)}\mathcal{O}_{S}\mathcal{A}^{(+)}_{z(0)}(x) \Bigg\}\delta(z-z_\alpha) \\
&=-\frac{1}{\sqrt{\pi R}}\int d^4x \Bigg\{\mathcal{N}^{(+)}_0 e^{-\frac{1}{3} k|z_\alpha-z_p|}\mathcal{J}_\mu \mathcal{A}^{\mu +}_{(0)}(x) + \mathcal{N}^{(+)}_0 e^{-\frac{1}{3} k|z_\alpha-z_p|}\mathcal{O}_{S}\mathcal{A}^{(+)}_{z(0)}(x)\\ 
&+ \sum_{n=1}^{\infty}\mathcal{N}_{n}^{(+)}\Bigg[-\frac{k R}{6 n} \sin\left[\frac{2n}{R} |z_\alpha-z_p|\right] + \cos\left[\frac{2n}{R} (z_\alpha-z_p)\right]\Bigg]\mathcal{J}_\mu \mathcal{A}^{\mu +}_{(n)}(x) \Bigg\},
\end{split}
\end{equation}
Analogous to the scalar case, and going by the argument
  mentioned therein, the dilaton factor $e^{-7S(z)/6}$
  has been introduced to induce localisation of the massless modes (both the vector and the scalar) alone
towards one of the branes. Further, from eq.\ref{eqn:veceve} -
\ref{eqn:vecodd}, we see that the coupling profiles of the KK modes
at the branes is of the same nature as
obtained for the scalar theory and reflects the \textit{clockworking}
attribute of the corresponding discrete theory.

By analogy with the scalar case, again, it is straightforward to see
that the EOMs in eq.\ref{eqn:eom_vector} have solutions corresponding
to asymmetric BCs as well with no mass degeneracy. Of course, in this
case, there exists a light vector mode with a nonzero mass, with the
corresponding light scalar acting as a Goldstone boson which vanishes
altogether in the unitary gauge.  This can also be understood by
noting that the \textbf{(DN)} BC does not preserve the 4D gauge
invariance of the zeroth mode as opposed to that in the \textbf{(NN)}
case.

\section{Perturbation and fine tunings in the LD theory: An aside} \label{sec:ft}
The degeneracy between the odd- and
even-modes could be considered a surprise, and it is worth pondering
if it is the result of a fine-tuning of parameters. Given this, it is worth reexamining, at ths stage, the linear dilaton theory of
Sec.\ref{sec:contgeom} for the extent of any extra fine-tuning beyond
the usual one encountered in generic warped theories associated with
the vanishing of the 4D cosmological constant (CC). Apart from the
equality between the brane tensions and the bulk curvature
mentioned in Sec.\ref{subsec:cwgeom}, {\em viz.}
\beq
\Lambda_2=4 k M_5^3=-\Lambda_1=-\Lambda_3\ ,
\label{brane_tension_tuning}
\eeq
the present theory seemingly
demands an additional tuning in order to arrange for the relation
$\mathcal{V}_1=\mathcal{V}_3$ of the VEVs required to stabilize the
fifth dimension.  However, note that, barring the last equality in
eq.(\ref{brane_tension_tuning}), the rest is exactly the same as in
the usual CW (or its linear-dilaton cousin) and does not introduce any
fine-tuning beyond the usual one. As for the last equality, clearly
$\Lambda_1 = \Lambda_3$ is mandated by the ${\mathbb Z}_2$ inherent to
the formulation and hence, unless the ${\mathbb Z}_2$ is explicitly
broken, this does not represent an extra fine-tuning
either. Exactly same is the case for
  $\mathcal{V}_1=\mathcal{V}_3$.  At this level, then, the degeneracy
between the even and odd states of a bulk field in the LD background
seems a natural outcome.

The situation, however, could change drastically if we relax either of
the two conditions inherent to the said degeneracy, namely the
${\mathbb Z}_2$ and the uniformity of the CW coupling (reflected, in
the continuum, by a uniform warping). We now examine each in turn,
starting with the second. While our treatment might (and justifiably)
be termed an ad hoc one, in the absence of any knowledge of the
fundamental theory that stipulates either the ${\mathbb Z}_2$ or the
properties of the dilaton, possible deviations from our assumptions
can hardly be dismissed, and, indeed, could very well be caused by
quantum corrections, if nothing else. We maintain that the discussion
below, while only an illustrative one, yet captures the qualitative
features.

\subsection{Non-uniform warping}

  We begin by examining the case where the ${\mathbb Z}_2$ is intact,
  but the warping is no longer uniform (linear). In other words, we
  are interested in examining the consequences of replacing
\[
\sigma(z) = \frac{-2}{3} k |z - z_p|
\to \frac{-2}{3} K(z) |z - z_p|
\]
where $K(z)$ varies from a constant value by only a small amount, {\em viz.}
\beq   K(z) = k + \delta k(z) \ ,
\label{k-change}
\eeq
so that a perturbative treatment may be permissible.  As argued above,
such a change may be brought about by, for example, altering the
dilaton potential (presumably away from the linear dilaton paradigm)
and adjusting the brane potentials adequately so as to have a
consistent solution.  However, rather than delve into the origin of
such a change, we treat eq.(\ref{k-change}) as determining an
effective background for the bulk scalar field $\phi$.

Such a change would, naturally, be expected to break the degeneracy
between the even and odd modes by an extent determined by the
magnitude of the shift in $k$. With $\delta k(z)$ being an unknown
function, admitting even oscillatory behaviour, a robust measure of
this shift is necessary and, to this end, we define a relative shift
magnitude by
\beq
\frac{\delta_K}{k} \equiv \left(
\frac{1}{\pi R} \int_0^{\pi R} dz \left[\frac{\delta k(z)}{k}\right]^2
\right)^{1/2} \ .
  \label{delta_k_meas}
\eeq
A nonzero $\delta k(z)$ would lead to a shift in the hitherto
degenerate (for each $n$) mass-squared matrices, not only in the
diagonal terms, but also in the off-diagonal ones.  For small
$\delta_K$, these shifts can be calculated perturbatively, leading to
\beq
\begin{split}
    \Delta m_n^{(++) 2} &= \frac{2k}{\pi R}\int_{0}^{\pi R} dz \, e^{3 \sigma(z)}\, \left[f_n^+(z)\right]^{2}\, \delta k(z) \\
    \Delta m_n^{(--) 2} &= \frac{2k}{\pi R}\int_{0}^{\pi R} dz \, e^{3 \sigma(z)}\, \left[f_n^{-}(z)\right]^2 \, \delta k(z) \\
    \Delta m_n^{(+-) 2} &= \frac{2k}{\pi R}\int_{0}^{\pi R} dz \, e^{3 \sigma(z)}\, f_n^{+}(z)f_n^{-}(z)\, \delta k(z)
\end{split}.
\eeq
(Of course, if ${\mathbb Z}_2$ is maintained by $\delta k(z)$, then
$\Delta m_n^{(+-) 2} = 0$). Given the above, the
mass-splitting introduced at each KK level is given by
\beq
\Delta m_n^2 = \sqrt{\left( \Delta m_n^{(++) 2}-\Delta m_n^{(--) 2}\right)^2+ 4\left(\Delta m_n^{(+-) 2}\right)^2},
\eeq
With the mass splitting being an observable, the functional dependence
of the relative splitting $\Delta
  m_n / m_n$ on the quantity $\delta_K / k$ can be used as a
diagnostic for the extent of fine tuning in the theory
\cite{Barbieri:1987fn}. To quantify this, we need to specify the form
of $\delta k(z)$. As illustrative examples, we consider two forms, namely
\beq
\begin{split}
    {\rm Case \, I:} \qquad \delta k(z)&= \epsilon \, k \cos{\left[k\left(z-\frac{\pi R}{2}\right)\right]} \\
    {\rm Case \, II:} \qquad \delta k(z)&= \epsilon \, k \exp{\left[-k\left|z-\frac{\pi R}{2}\right|\right]},
\end{split}
\eeq
where $\epsilon$ is a small constant. Fig.\ref{fig:tuning} shows the
consequent variation of the
relative splittings with the fractional  shift in $k$.
\begin{figure}[htb] 
\centering
      \includegraphics[scale=0.4,keepaspectratio=true]{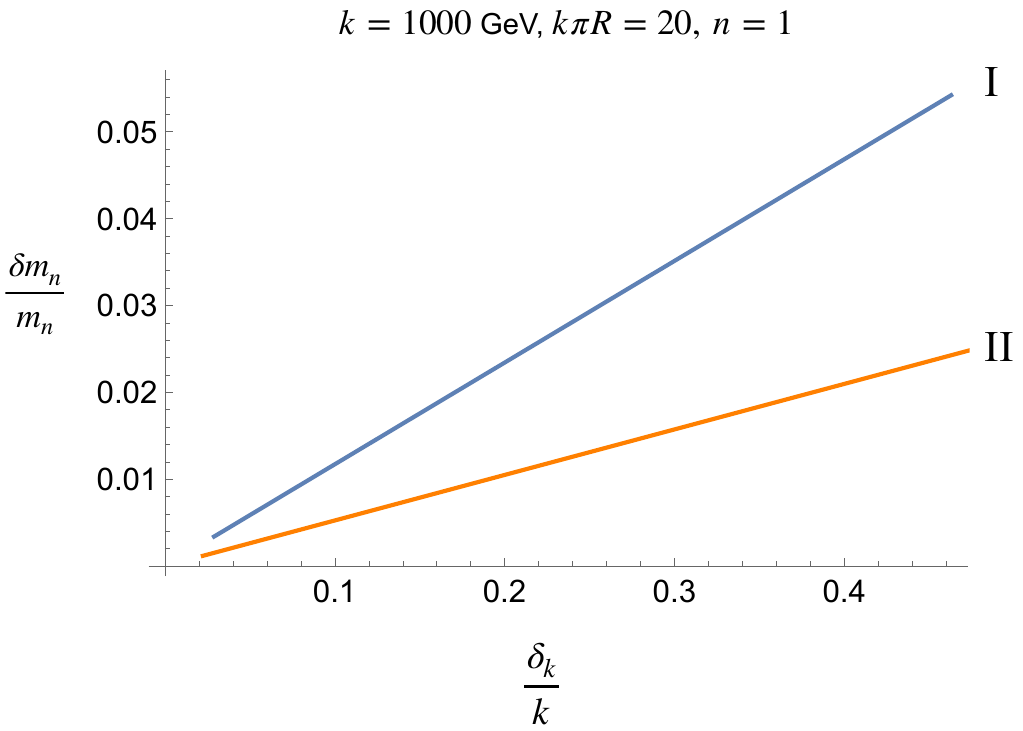}
      \includegraphics[scale=0.4,keepaspectratio=true]{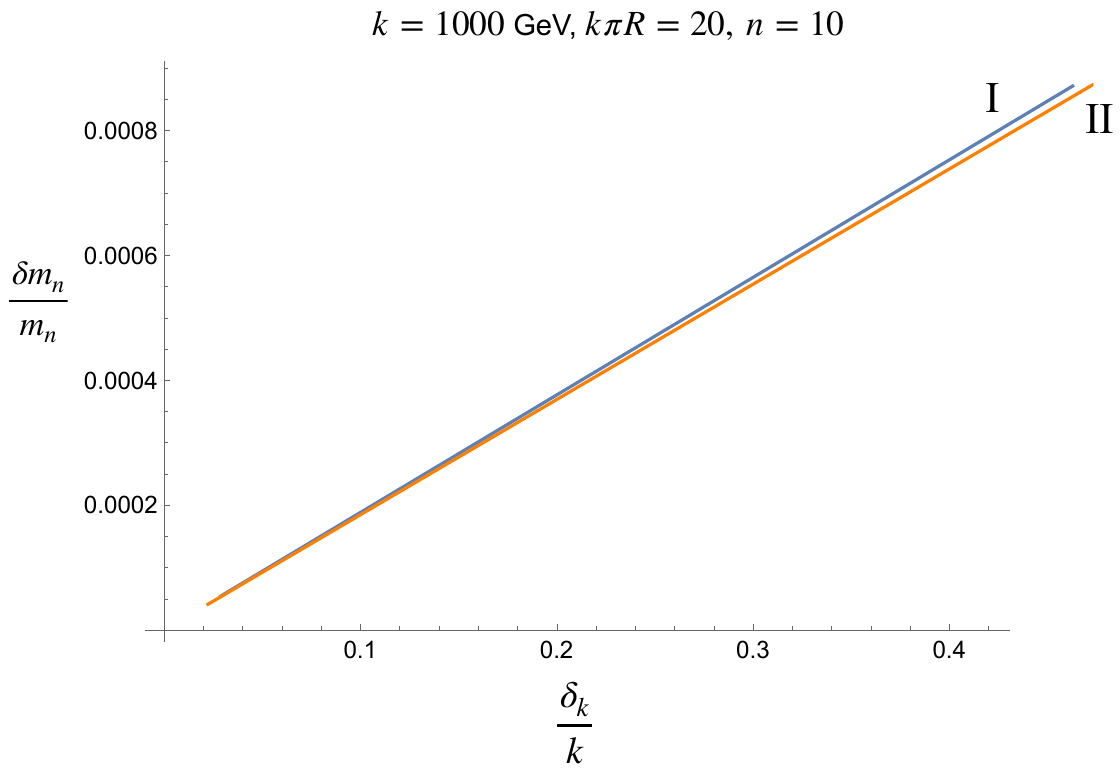}
      \caption{Relative mass splitting as a function of the fractional change $\delta_K / k$  for the KK levels (a) $n=1$ and (b) $n=10$.}
	 \label{fig:tuning}
\end{figure}
A few features are immediately clear:
\begin{itemize}
\item With the products $e^{3 \sigma(z)} \left[f_n^\pm(z)\right]^2$ being simple
    oscillatory functions for $n>0$, the
  extent of the correction is governed, to a very large extent, by the
  profile of $\delta k(z)$.
\item Typically, $\Delta m_n /m_n \ll \delta_K / k$. This is not
  difficult to understand, especially for the two profiles that is
  presented. Although, in general, the contribution to $\Delta
  m_n^{(++) 2}$ and $\Delta m_n^{(--) 2}$, individually, would tend to
  be maximized when the peaks of $\delta k(z)$ coincide with the peaks
  of $\exp{[3 \sigma(z)]} f^2$, the oscillatory behaviour of the
  latter would tend to average out the two contributions to
  comparable values. Thus, the consequent splitting would not change
  significantly for other profiles as well.
\item In particular, for Case I, the oscillator profile leads to cancellations
  between corrections from different $z$-regions. Even then, the corrections are
  still much larger than those for Case II, wherein 
  $\delta k(z)$ quickly falls off away from the central brane leading to
  sizable contributions only from a narrow sliver.
\item The absolute splittings tend to be smaller for higher KK-levels as the weight
  function $\exp{[3 \sigma(z)]} f^2$ becomes rapidly oscillating as
  $n$ increases, leading to $\Delta m_n^{(++) 2}$ and $\Delta
  m_n^{(--) 2}$ being similar in magnitude. The relative
    splitting is further suppressed as the (unaltered) contribution
    from the momentum in the fifth direction grows quickly.
\end{itemize}
In summary, the relative smallness of $\Delta m_n /m_n$ {\em vis a vis}
the `perturbation' $\delta_K / k$ implies that the Barbieri-Giudice~\cite{Barbieri:1987fn}
measure of fine-tuning is quite small in the present context.

\subsection{A broken ${\mathbb Z}_2$}

We, now, turn to the second possibility mentioned above, namely
maintain a linear $\sigma(z)$, but break the discrete symmetry.
Naively, the simplest way to achieve this would be to either shift the
intermediate brane from the geometrical centre or assume the mass
parameter $k$ to be asymmetric around the central fixed point in the
bulk, or both.

It should be realized, though, that consistency would, generically
demand that unequal $k$ in the two halves should be accompanied by a
shift in the position of the central brane. The easiest way to
understand this is to fall back on the discrete case, where $q$---the
charge scaling factor between consecutive sites---had taken the place
of $k$ in the continuum and $p$---the number of steps between the two
pivots---was analogous to the radius $ r \equiv R/2$.  If the pair
$(q,p)$ were different in the two halves, clearly, consistency at the
two pivots demands\footnote{The ensuing arguments are qualitatively
  valid for $\mathbb{Z}_2$ asymmetric $m$ and $f$ parameters as well
  in the discrete theory.} $q_1^{p_1} = q_2^{p_2}$. In the continuum
language, this translates to
  \beq \label{eq:tunr12}
  \begin{split}
     r_1 = R-r_2=\frac{k_2}{k_1 + k_2}R \ ,
  \end{split}
  \eeq
where $r_{1,2}$ denote the radii of the patches to the left and right
of the intermediate brane, respectively.  Explicitly, for $z_p \neq
\pi R/2$ and unequal values of the parameter $k$ about $z_p$, we
obtain the solution
\beq
  \sigma(z)=
  \left\{
  \barr{rcl}
\dis \alpha_1 \frac{\Lambda_2}{3 M_5^3} (z-z_p) &  \qquad z < z_p, \\[3ex]
\dis -\alpha_2 \frac{\Lambda_2}{3 M_5^3} (z-z_p) &  \qquad z > z_p
\earr
\right.
\eeq
with the 4D CC tuning now amounting to $\Lambda_2=2 M_5^3 (k_1 +
k_2)=-\Lambda_1 / 2 \alpha_1 =-\Lambda_3 / 2 \alpha_2 $. Here, $k_1$
and $k_2$ denote the constant\footnote{Field equations of the LD
  theory only allow for flat variations of $k$ in the two patches
  around the central fixed point. A generic variation $\delta k(z)$
  would, perhaps, require venturing beyond the linear dilaton paradigm
  and introducing additional degrees of freedom in the bulk.} mass
parameters for $z<z_p$ and $z > z_p$, respectively, and $\alpha_{1,2}$
are defined as
\beq
\alpha_1 \equiv \frac{2 k_1 + k_2}{3(k_1 + k_2)}  \quad {\rm and} \quad
\alpha_2 \equiv \frac{k_1 + 2 k_2}{3(k_1 + k_2)} \ .
\eeq
In other words, we have
\beq
  \sigma(z)=
  \left\{
  \barr{rcl}
\dis \frac{2}{9} (2 k_1 + k_2) (z-z_p) &  \qquad z < z_p, \\[3ex]
\dis \frac{-2}{9} (k_1 + 2 k_2) (z-z_p) &  \qquad z > z_p \ .
\earr
\right.
\eeq
The relation which the VEVs must satisfy for radius stabilization now
becomes,
\beq
\mathcal{V}_1 - \mathcal{V}_3
= \frac{2\pi}{3} \left[(k_1 + 2 k_2) r_2 - (2 k_1 + k_2) r_1\right]
=\frac{2\pi}{3}(k_1 - k_2)R \ .
\eeq
where, the last equality is obtained using the relation in eq.(\ref{eq:tunr12}).
The equality of the potentials on the brane, {\em viz.}
  $\mathcal{V}_1=\mathcal{V}_3$, therefore, can be
understood as being tantamount to a fine-tuning required to preserve
the $\mathbb{Z}_2$ parity.

It should be remembered that, with the ${\mathbb Z}_2$ now having been
explicitly broken, there is no notion of even and odd solutions, far
less mass degeneracy. Indeed, the only permissible nontrivial
solutions to the equation of motion (with similar forms of the BCs as
assumed originally) are those that have support on only one of the two
domains, left or right. This is not difficult to understand as, on
breaking the ${\mathbb Z}_2$, all that we have are two contiguous, yet
disparate patches, that have been sewed together. In particular, the
solutions {\em have to vanish} (namely a Dirichlet boundary condition)
at the boundary between the two patches. The number of independent
solutions remains the same though, the erstwhile even and odd pairs
being replaced by the pair of solutions over the individual patches.

The mass splitting, in this case, is easy to ascertain, being given
by the expressions in eq.(\ref{eqn:scaleven}), but computed
with differing $k_{1,2}$.

\subsection{The generic case}
The most general situation (where both the ${\mathbb Z}_2$ is broken
and a complicated warping is in place) can be qualitatively understood
in terms of the two simplified discussions above. The Barbieri-Giudice
measure for fine-tuning is expected to remain tolerable (as long as it
is not applied to the isuue of the vanishing cosmological constant, in
common to all other such warped models). Furthermore, the features
alluded to above would continue to remain applicable.

\section{Summary} \label{sec:conclu}
The clockwork mechanism provides a novel way of generating
hierarchical couplings without resorting to large fine tunings. The
fact that the original class of clockwork theories also finds a
correspondence with the 5D linear dilaton model makes it a compelling
avenue to be explored further. In this paper, we substitute the
original open ended construction by a closed chain structure of the
near-neighbour interactions. The resulting $\mathbb{Z}_2$ symmetry
under field exchanges about the centre-most site gives rise to two
sets of physical states, one even and the other odd under an exchange
parity. The even massless state is localised towards one of the
$\mathbb{Z}_2$ fixed points (the pivot sites) reflecting the clockwork
mechanism. Keeping this symmetry intact then renders the lightest odd
state absolutely stable and, hence, a potential dark matter candidate.

This new class of theories can also be obtained as a deconstruction of a bulk 5D theory in a linear
dilaton background invariant under a $\mathbb{Z}_2$
transformation about a fixed point identified
with the centre of the compact extra
dimension. The discretization prescription, however, needs to be
amended so as to preserve the
$\mathbb{Z}_2$ symmetry. The underlying $\mathbb{Z}_2$
invariant warped background metric is obtained as the {\em
exact} ({\em i.e.}, including backreaction) solution of an extended linear dilaton
theory on an $S^1/\mathbb{Z}_2$ orbifold with three equidistant
3-branes.
The theory, therefore, potentially provides two
physically inequivalent setups, namely, the IR-UV-IR and the UV-IR-UV
scenarios, both admitting different, yet
viable solutions to the EW-Planck hierarchy problem.

Bulk field theories (whether scalar or vector) in this background, on
compactification, are characterised by a physical
spectrum consisting of two KK towers with states
that are even and odd under the $\mathbb{Z}_2$ parity, mimicking the notion of KK parity in typical UED models
with a flat metric.  With identical boundary conditions at the
orbifold fixed points (NN and DD), the theory admits a massless mode
and the two KK towers (beyond a mass gap $\sim k$) turn out to be
doubly degenerate at each level with mass eigenvalues that
match those in the discrete theory in the large $N$ limit
at leading order in $(1/N)$. This scenario, therefore, elucidates the
correspondence between the discrete and the continuum clockwork
theories. On the other hand, with non-identical boundary conditions,
we obtain a non-degenerate spectrum of heavy KK states along with a
light mode that has a small, albeit nonzero, mass. While this
picture does not immediately relate to the
discrete theories that we introduced, it is potentially interesting in its own
right. In particular, it demonstrates that
the choice of boundary conditions is germane to the
aspect of establishing a correspondence between a discrete CW theory
and a bulk 5D theory in the linear dilaton background.

In short, this study demonstrates that the clockwork mechanism and its
correspondence with the linear dilaton model need not be associated
with only a specific topology of the discrete and continuum theories
and that extensions within this CW/LD framework are possible, like the
one we have discussed in this work which projects a richer
phenomenology and, presumably, new and interesting model building
applications.

\appendix
\section{Mass Degeneracy} \label{sec:deg}
\subsection{Degeneracy in the continuum CW} \label{subsec:condeg}
Reexamining the equation of motion
(eq.\ref{eqn:bscaleom}) for the bulk modes of the scalar theory, namely,
\begin{equation}
\left(\partial_z^2 -k^2 + m_n^2 \right)\left(e^{3\sigma(z)/2 }f_n(z)\right)
= 0 \ ,
\end{equation}
we immediately see that, if $f_n(z)$ satisfies the equation, so does
$f'_n(z)$ and for the same $m_n$. In other words, the two
wavefunctions, if admissible, are degenerate, and if they are
$\mathbb{Z}_2$ eigenstates, they must correspond to opposite
eigenvalues.  It is easy to verify that the odd \textbf{(DD)}
solutions are indeed derivatives of the even \textbf{(NN)} solutions,
{\em viz.}
\beq \label{eqn:degrel}
f^{(-)}_n(z)=-m_n^{-1}\partial_z f^{(+)}_n(z).
\eeq
Naively, it might seem that further derivatives would produce more
candidate eigenstates for a given $m_n$. However, note that the
boundary conditions restrict the admissible solutions. As for the
remaining, these can be written as linear combinations of the
(fundamental) solutions $f^{(\pm)}_n$ for each KK level $n$ and,
hence, are not independent. This is expected as the solutions
$f^{(\pm)}_n$ span the two-dimensional degenerate space that is stipulated by the
$\mathbb{Z}_2$ symmetry. Clearly, the
same argument also holds for the vector field theory.

In contrast, for the mixed BC case (\textbf{DN} and \textbf{ND}), the
BCs at $z=0$ corresponding to the even and odd modes of the same KK
level $n$ are incompatible with a relation like eq.\ref{eqn:degrel}
which serves to explain the non-degeneracy encountered in that case.

We end this discussion by reexamining the source of this
degeneracy. Consider any deformation along the $z$-dimension, {\em
e.g.}, a different value of the parameter $k$ over a patch in the bulk
(and, to ensure $\mathbb{Z}_2$ symmetry, a corresponding patch in the
opposite region). This would introduce continuity conditions on the
derivatives of the solutions that, {\em e.g.}, are $\mathbb{Z}_2$-odd
for even solutions. The mass eigenvalues would have a dependence on
these derivative conditions, thereby, breaking the degeneracy between
the even and the odd modes. Thus, it is the $\mathbb{Z}_2$ symmetry,
the boundary conditions and the universality of the parameters in the
theory that, {\em together}, facilitate the emergence of a degeneracy
in the physical mass spectrum.

\subsection{Degeneracy in discrete CW} \label{subsec:disdeg}
While the degeneracy in this case could be motivated from that
for the continuum, we examine the issue on its own.
As previously mentioned, the degeneracy in the spectrum emerges by
virtue of the three characteristic features of the theory, namely, the
universality of the parameters along the sites, the $\mathbb{Z}_2$
symmetry and the specific form of interactions involving the pivot
sites. It is, perhaps, easier to trace the consequences of each of the
properties individually in order to verify the preceding
assertion. To start with,
the universal nature of the parameters essentially
leads to a (pseudo-)tridiagonal mass matrix with deviations only in the
$0$-th, $p$-th and $(2p-1)$-th rows (which, for brevity, we address as
the \textit{pivot} rows) with these being
a consequence of the $\phi_0-\phi_{2p-1}$ and the
$\phi_{p-1}-\phi_{p}$ mixings.  The resulting eigenvalue equations,
barring those involving
the pivot ones, are second order linear difference equations
of the form
\beq
-q a_{j-1,n}+(1+q^2-\lambda_n)a_{j,n} -q a_{j+1,n}= 0,
\eeq
with equal coefficients for $a_{j-1,n}$ and $a_{j+1,n}$. Here,
$\lambda_n$ denotes the $n$-th eigenvalue. This equation is nothing
but a second-order $q$-difference equation or, in other words, the
discretized version of the harmonic equation with ordinary derivatives
replaced by $q$-derivatives. Consequently, the solutions (away from
the pivot points) are given by
\beq \label{eq:deggen}
a_{n j}=\mathcal{N}_n\begin{cases}
\sin{j\theta_n}+B\cos{j\theta_n}, & j\leq p \\
C\sin{j\theta_n}+D\cos{j\theta_n}, & j> p
\end{cases},
\eeq
where $B$, $C$ and $D$ are constant coefficients, $\mathcal{N}_n$ are
the normalisation factors and $\theta_n$ is
an {\em a priori} undetermined function of
$n$. Reverting to the eigenvalue equation for the pivot rows, we have
\beq \label{eq:piveqn}
\begin{split}
    2q^2 a_{n,0}-q a_{n ,1} -q a_{n, N-1}&=\lambda_n a_{n,0} \\
    -q a_{n ,p-1} + 2 a_{n, p} -q a_{n, p+1}&=\lambda_n a_{n,p} \ .
\end{split}
\eeq
Furthermore, the eigenvalues have the form,
\beq \label{eq:degeval}
\lambda_n = 1+q^2-2q\cos{\theta_n}. 
\eeq
Thus, in effect, the universal parametrisation fixes the form of the
eigensystem to the simple expressions given in eq.\ref{eq:deggen}
and \ref{eq:degeval}.

Applealing, now, to the $\mathbb{Z}_2$ symmetry, it stipulates the existence of even and odd solutions. Consequently, one obtains for the odd eigenvectors
$a^-_{k,p}=0$ which implies $B=-\tan{p\theta_n}$.

Finally, with the
odd solutions satisfying the second equation in \ref{eq:piveqn}
trivially, we see that the first pivot equation determines the exact
form of the eigenvalues by specifying $\theta_n = n \pi / p$, or,
alternatively,
\beq
\lambda_n = 1+q^2-2q\cos{\frac{n \pi}{p}},
\eeq
which clearly are the doubly-degenerate eigenvalues discussed before
in sec.\ref{subsec:discscal}. The pivot equations, therefore, play the
role of boundary conditions about the $\mathbb{Z}_2$ fixed points as
encountered in the continuum picture. In summary, the preceding
arguments establish that the structural properties of the theory
intricately conspire to give rise to the degenerate mass spectrum and
that a deformity in any of the three properties would break the
degeneracy.

\section{Dynamical dilaton modes and IR-UV-IR/UV-IR-UV} \label{sec:dyndil}
In sec.\ref{subsec:hier} we had discussed
briefly the emergence of two kinds of scenarios in our triple brane
theory, namely, the IR-UV-IR and the UV-IR-UV setups\footnote{We
specify an IR brane as one where any localised theory, as viewed in
the effective 4D metric, would have its physical mass parameters set
by the TeV scale. Similarly a UV brane is one where a theory would
have all its mass parameters, ostensibly, near the Planck
scale. Hence, for $k>0$ the IR branes are situated at the boundaries
of the orbifold with the UV brane located in the middle, and the other
way around for $k<0$.}, corresponding to $k>0$ and $k<0$
respectively. Each of these provides
possible explanations of the hierarchy problem. Here, we delineate the
stability of the two cases with regards to the physicality of the
dynamical modes generated therein. Concentrating on scalars, the said modes
can be identified by
considering, at the linear order, the scalar
fluctuations\footnote{Note that the vector fluctuations of the metric,
alongwith one combination of the two scalar flucuations, 
disappear in the unitary gauge to resurface as the longitudinal modes of the
massive tensor modes.} $\Psi(x,z)$ and $\Phi(x,z)$
in our CW metric \cite{Kofman:2004tk}, {\em viz.}
\beq \label{eqn:scalpert}
ds^2=F(z)^2\left[(1+2\Psi(x,z))\eta_{\mu \nu} dx^\mu dx^\nu + (1+\Phi(x,z))\right], \qquad F(z)\equiv e^{\sigma(z)}.
\eeq
Furthermore, the perturbation of the linear dilaton
field $S$ in eq.\ref{eqn:ldgaction} is denoted by $\delta
S(x,z)$. On linearization of the equations of motion, the aforementioned separation of the scalar mode from the others is encapsulated in
 the constraint equations
\beq \label{eq:consteq}
\begin{split}
    &\Phi-\frac{F}{F'}\Psi'-\frac{F}{9 F'}S'\delta S=0 \\
    &\Phi=-2\Psi
\end{split},
\eeq
where the primes denote $z$-derivatives. On using these, only one independent bulk scalar remains, and its
dynamical equation reduces to
\beq
\left( \partial_x^2 + \partial_z^2-k^2\right)\left[e^{3\sigma(z)/2} \Phi(x,z) \right]=0.
\eeq

On effecting the decomposition
\beq
\Phi(x,z)=\sum_n Q_n (x) \tilde{\Phi}_n(z) \ ,
\eeq
 equations of motion are decoupled, and
\beq
\begin{split} \label{eqn:appeom}
&\left(\partial_x^2 - m_n^2 \right)Q_n(x)=0 \\
&\left(\partial_z^2-k^2+m_n^2\right)\left[e^{3\sigma(z)/2} \tilde{\Phi}(z)\right]=0.
\end{split}
\eeq
Additionally, by integrating the linearised Einstein
equations (including the boundary terms) over
a small interval about the fixed points, we obtain two junction
conditions (JC) at each brane for the perturbations. One of these is
equivalent to eq.\ref{eq:consteq} when evaluated at the
branes and, hence, imposes no new constraints. The remaining
condition, namely,
\beq
\delta S'-S'\Phi \Bigg|^{z_\alpha+\epsilon}_{z_\alpha-\epsilon}=6 e^{\sigma(z)}\lambda_\alpha''(S)\delta S \Bigg|_{z_\alpha},
\eeq
relates the perturbation $\delta S$ to the parameters at the
branes. This, in turn, can be substituted in
eq.\ref{eq:consteq} to obtain JCs operative directly on $\Phi$,
one at each fixed point\footnote{Note that on account of the
$\mathbb{Z}_2$ symmetric structure of the theory about the central
fixed point, the junction condition at one of the orbifold boundaries
trivially fixes that on the opposite boundary.}. Here,
$\lambda_\alpha(S)$ are the brane potentials, as defined in
eq.\ref{eq:brane_pt} and the primes on $\lambda_\alpha$ denote
differentiations with respect to $S$.  The JC at the $z=z_p$ brane
generates the solutions
\beq
\barr{rcl}
\dis \tilde{\Phi}_0(z)& \propto& \dis
e^{-3\sigma/2}\left \{ \sinh{\left[ \tilde{\beta} |z-z_p|\right]}+\tilde{\gamma} \cosh{\left[ \tilde{\beta} (z-z_p)\right]}\right\}
\\[2ex]
\dis \tilde{\Phi}_{n>0}(z)& \propto & \dis
e^{-3\sigma/2}\left \{ \sin{\left[ \beta_n |z-z_p|\right]}+\gamma_n \cos{\left[ \beta_n (z-z_p)\right]}\right\},
\earr
\eeq
where $n=0$ corresponds to the radion mode and
\beq
\barr{rclcrcl}
\dis \tilde{\beta}^2 &\equiv & \dis k^2-m_0^2 &\qquad &
\dis \tilde{\gamma} & \equiv & \dis
\frac{3 \tilde{\beta} \mu_p}{(k \mu_p-k^2+\tilde{\beta}^2)}
\\[2ex]
\dis \beta_n^2 &\equiv &\dis m_n^2-k^2  & &
\gamma_n &\equiv & \dis \frac{3 \beta_n \mu_p}{(k \mu_p-k^2-\beta_n^2)} \ .
\earr
\eeq
Here, $\mu_\alpha$ are the mass parameters in the brane potentials
$\lambda_\alpha(S)$, as detailed in
eq.\ref{eq:brane_pt}. Solving
the equations emerging from the JC at $z=0$, on the other hand,
specifies the mass spectrum which, in the limit
$\epsilon_\alpha \equiv |k|/\mu_\alpha \ll 1$ and upto linear order in
$\epsilon_\alpha$, are given by

\beq
\barr{rcl}
m_0^2 & \simeq & \dis
\left\{
\barr{rclcl}
\dis \frac{8k^2}{9}\left( 1-\frac{1}{9}\left[ (\epsilon_0 - \epsilon_p)+(\epsilon_0 + \epsilon_p)\coth{\frac{k\pi R}{6}}\right]\right)
&\simeq & \dis
\frac{8 k^2}{9}\left(1-\frac{2 \epsilon_0}{9}\right) & \quad & k>0
\\[3ex]
\dis \frac{8k^2}{9}\left( 1-\frac{1}{9}\left[ (\epsilon_p - \epsilon_0)+(\epsilon_0 + \epsilon_p)\coth{\frac{k\pi R}{6}}\right]\right)
& \simeq & \dis \frac{8 k^2}{9}\left(1-\frac{2 \epsilon_p}{9}\right) && k<0
\earr
\right.
\\[7ex]
m_n^2 &\simeq & \dis
k^2 + \frac{4 n^2}{R^2}\left[1-\frac{12(4n^2+k^2 R^2)}{|k|\pi R(36 n^2+k^2 R^2)}(\epsilon_0 + \epsilon_p) \right]\ .
\earr
\eeq

At the second order in the scalar perturbations the dynamical field, after diagonalisation of the action, is specified by a combination of $\Phi$ and $\delta S$, namely, 
\beq
\xi(x,z)\equiv \left( \frac{\Phi}{2}-\frac{\delta S}{3}\right)= \sum_n Q_n (x) \tilde{\xi}_n(z),
\eeq
which satisfies the dynamical equation given in
eq.\ref{eqn:appeom}.

The free 4D actions for the physical modes are, then, given by
\beq
S_n =\mathcal{C}_n \int d^4x \, Q_n\left( \partial_x^2-m_n^2\right)Q_n \, ,
\eeq
where
\beq
\mathcal{C}_n \equiv \frac{3 M_5^3}{2}\int_0^{\pi R} dz \, F^{-1}\left[ \frac{9}{2 S'^2}\left( F^2 \tilde{\Phi}_n\right)'^2 + \left(F^2 \tilde{\Phi}_n \right)^2 \right] > 0.
\eeq

The positivity of $\mathcal{C}_n$ 
ensures that the theory is free of negative kinetic terms (and, hence, ghosts)
for either sign of $k$. It is easy to ascertain that, for,
$\epsilon_\alpha \ll 1$, the theory is free of tachyonic modes (as exemplified by a negative mass-squared) as
well. For large values of $\epsilon_\alpha$, however, one needs to
perform the stability analysis numerically as closed-form
analytical expressions
for the KK masses are not straightforward.
Moreover, invoking
non-minimalities, such as assuming a Higgs-curvature interaction on a
brane, could lead to a ghost radion field depending on the value of
the mixing parameter \cite{Cox:2012ee}. In the spirit of the RS
gravity models \cite{Davoudiasl:2003zt, George:2011sw}, adding
brane-localised curvature terms could potentially introduce a ghost
radion as well and, hence, would constrain their coefficients. Thus,
in such cases, the stability of the two scenarios ought to be inspected
again thoroughly. It should also be borne in mind that a possible
embedding of linear dilaton models in a string theoretic
setting \cite{Antoniadis:2011qw} offers a way to ameliorate putative instabilities arising on account of
negative tension branes.

\acknowledgments

S.M. acknowledges research Grant
No. CRG/2018/004889 of the SERB, India.
\bibliography{ccw_revised.bib}
\end{document}